\pdfoutput=1
\PassOptionsToPackage{unicode}{hyperref}
\PassOptionsToPackage{hyphens}{url}
\documentclass[
]{article}
\usepackage{amsmath,amssymb}
\usepackage{lmodern}
\usepackage[T1]{fontenc}
\usepackage[utf8]{inputenc}
\usepackage{textcomp} % provide euro and other symbols
\IfFileExists{upquote.sty}{\usepackage{upquote}}{}
\IfFileExists{microtype.sty}{% use microtype if available
  \usepackage[]{microtype}
  \UseMicrotypeSet[protrusion]{basicmath} % disable protrusion for tt fonts
}{}
\makeatletter
\@ifundefined{KOMAClassName}{% if non-KOMA class
  \IfFileExists{parskip.sty}{%
    \usepackage{parskip}
  }{% else
    \setlength{\parindent}{0pt}
    \setlength{\parskip}{6pt plus 2pt minus 1pt}}
}{% if KOMA class
  \KOMAoptions{parskip=half}}
\makeatother
\usepackage{xcolor}
\IfFileExists{xurl.sty}{\usepackage{xurl}}{} % add URL line breaks if available
\IfFileExists{bookmark.sty}{\usepackage{bookmark}}{\usepackage{hyperref}}
\hypersetup{
  hidelinks,
  pdfcreator={LaTeX via pandoc}}
\usepackage[margin=1in]{geometry}
\usepackage{longtable,booktabs,array}
\usepackage{calc} % for calculating minipage widths
\usepackage{etoolbox}
\makeatletter
\patchcmd\longtable{\par}{\if@noskipsec\mbox{}\fi\par}{}{}
\makeatother
\IfFileExists{footnotehyper.sty}{\usepackage{footnotehyper}}{\usepackage{footnote}}
\makesavenoteenv{longtable}
\usepackage{graphicx}
\makeatletter
\def\maxwidth{\ifdim\Gin@nat@width>\linewidth\linewidth\else\Gin@nat@width\fi}
\def\maxheight{\ifdim\Gin@nat@height>\textheight\textheight\else\Gin@nat@height\fi}
\makeatother
\setkeys{Gin}{width=\maxwidth,height=\maxheight,keepaspectratio}
\makeatletter
\def\fps@figure{htbp}
\makeatother
\providecommand{\tightlist}{%
  \setlength{\itemsep}{0pt}\setlength{\parskip}{0pt}}
\author{}
\date{}

\begin{document}

\hypertarget{comparing-measurements-of-pictorial-depth-conditions-for-placing-methods-observers-and-depiction-styles-on-one-scale}{%
\section{Comparing Measurements of Pictorial Depth: Conditions for
Placing Methods, Observers and Depiction Styles on One
Scale}\label{comparing-measurements-of-pictorial-depth-conditions-for-placing-methods-observers-and-depiction-styles-on-one-scale}}

\textbf{Adam Y. Shavit}\\
ORCID 0009-0008-1235-0995

\hypertarget{abstract}{%
\subsection{Abstract}\label{abstract}}

Two observers can agree about what lies in front in a picture yet report
different amounts of apparent depth. People report pictorial depth by
judging which of two points looks nearer, setting a probe to match a
surface, adjusting relative sizes, or rating three-dimensionality. Two
such methods can agree closely without sharing a unit. We ask when
different methods, observers, depiction styles and viewing conditions
can share one scale. Each set of scores, a measurement block, is a graph
node. Two blocks are linked only if their shared observations pass six
checks fixed in advance, among them outcome type, removal of declared
offsets and position effects, proportionality and covariance. If the
adjusted shared patterns are exactly proportional, a connected graph
fixes every block's gain, its relative scale factor. The checks accept
proportionality only within a set margin, so a link supports the gains
without proving exact proportionality. One unit is still not one depth
coordinate. Suppose each block scores shared stimuli with its own origin
and positive unit. A matrix-rank test then decides when a common
coordinate is identified: when only one overall origin and unit remain
free. If each shared stimulus, an anchor, is scored by exactly two
blocks, this needs two disjoint sets of anchors, each connecting every
block, their compatibility assumed, not tested. They also suffice,
except for anchor values satisfying exact polynomial relations
(probability zero for anchor values drawn from any distribution with a
density). We adapt network gain estimation from test linking, with a
loop-consistency check. A reanalysis of published probe settings on 3-D
renderings passes one of 171 candidate depiction-style links through all
six checks, each judged by its own 95\% bound, unadjusted for the 171
pairs. Links between tasks remain untested; they need shared stimuli and
a joint uncertainty model.

\hypertarget{introduction}{%
\section{1 Introduction}\label{introduction}}

Two observers can agree about what lies in front in a picture yet report
different amounts of apparent depth. Seven observers who set relative
sizes in one drawing produced depth maps that correlated very highly
with one another, yet each observer's map carried its own scale factor,
and the factors differed markedly between observers (Wagemans, van
Doorn, and Koenderink, 2011, ``Pictorial depth probed through relative
sizes'', pp.~1002-1003). Depth reports also change with the task. When
Koenderink, van Doorn, Kappers, and Todd (2001) had the same four
observers measure the same photographed sculptures with four tasks, the
raw depth maps often disagreed until an offset, a depth shear and a
depth gain were removed. Agreement about what is nearer therefore does
not ensure agreement about how much depth a picture shows. The
difference matters whenever observers, tasks or depiction styles are
compared as though their numbers shared a unit.

The phrase ``pictorial depth'' covers several operationally different
outcomes. A pairwise judgment, which of two points looks nearer,
identifies order. An exocentric pointing task reconstructs a
configuration from directed relations between picture points. A
relative-size task defines depth differences through the size ratios an
observer sets. Ratings of volume or local three-dimensionality concern
still other constructs. These measurements can agree closely on order or
shape without sharing a unit. The paper asks when measurements of
pictorial depth can be placed on one scale. It answers with
identification theorems, simulations and one worked reanalysis of
published data (Section 4), not with new data or image analysis.

The argument uses three terms. A measurement block is one set of depth
scores from one method, observer or group, picture, depiction style,
viewing condition and processing pipeline, and it is a node of a graph
(Section 2). A gate is a check, fixed before any ratio is fitted, that
an overlap between two blocks must pass before it may link them; Figure
1 lists the six gates, and an overlap that passes all six becomes an
active edge. An anchor is a stimulus scored by two or more measurement
blocks, which ties their scales together; in a depth-probe task it is
one probed point of a picture. Theorem 1 uses active edges to answer the
weaker question, whether blocks share a relative unit. Lemma 1 and
Proposition 2 use anchors to answer the stronger one, whether blocks
also share one depth coordinate, origin included.

\includegraphics[width=\textwidth,height=7.4in]{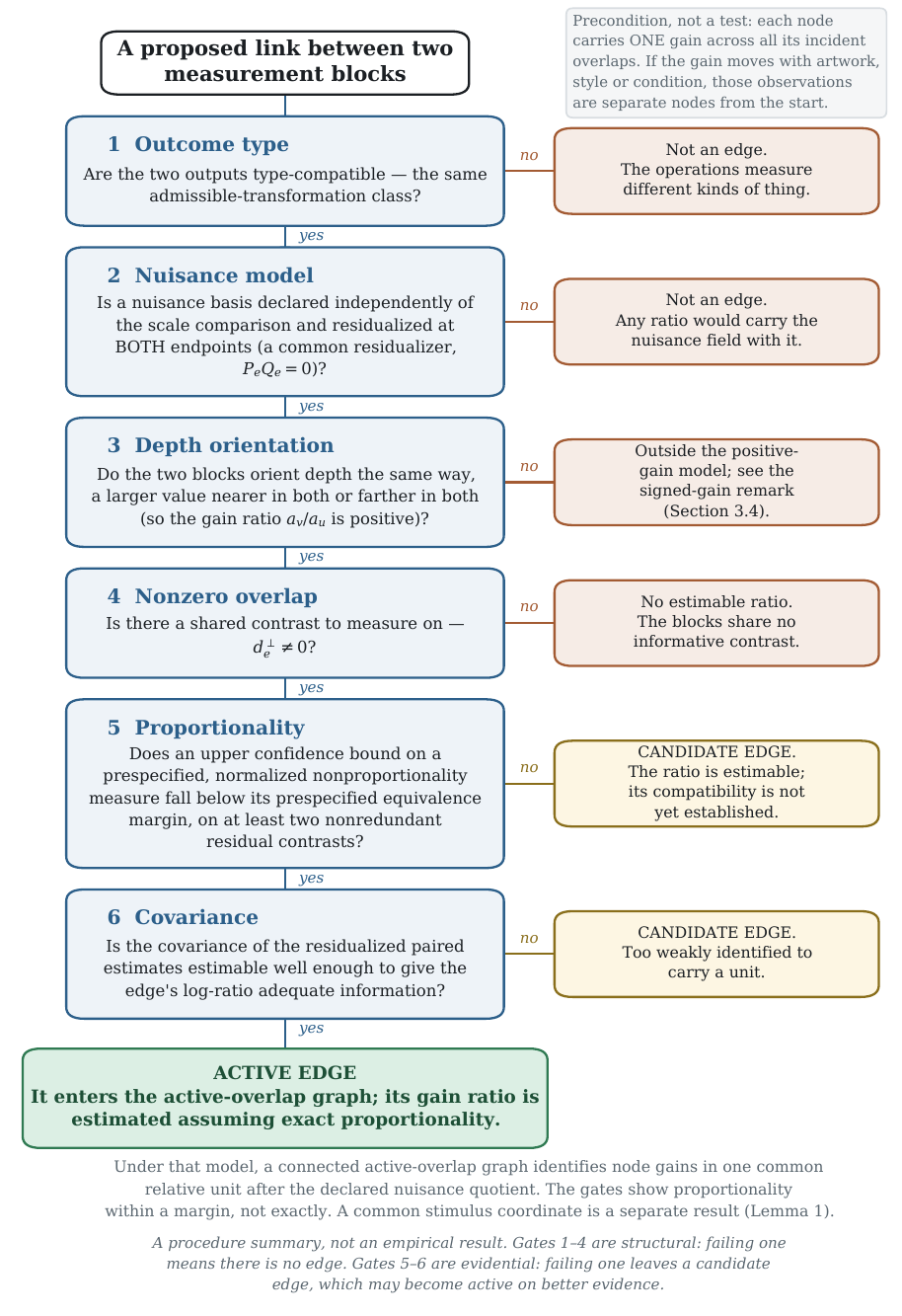}

Figure 1. Measurement-to-claim decision path. Six gates decide whether
an observed overlap may link two measurement blocks on a common relative
scale. The diagram is a procedure summary, not an empirical result.
Symbols are defined in Section 3.1. Gates 1--4 are structural: failing
one means there is no edge, although in data gates 3 and 4 are still
decided by a statistical test (Section 4). Gates 5--6 are evidential:
failing one leaves a candidate edge, which may become active on stronger
evidence. Passing gate 5 shows proportionality within a prespecified
margin, not exact proportionality, so an active edge enters the gain
estimate conditional on the exact model.

The paper's contributions are as follows. It gives a typed ontology for
pictorial-depth measurement operations. It sets gates that decide,
before any estimation, whether an overlap may link two blocks, judged
against a prespecified equivalence margin rather than a correlation. It
gives separate results for relative-unit linkage and for a common affine
representation. A one-edge variance score, an application of the
Sherman--Morrison identity, ranks candidate edges inside an already
connected part of the graph before they are run. Operational criteria
cover signal strength and deletion robustness. Simulations with frozen
seeds and specified generators separate algebraic identification from
precision. Section 3.6 gives a criterion for a design to stay identified
after losing any one anchor. Such a design needs at least one more
anchor than the minimum that bare identification needs. For both
properties the anchor count is necessary, and the arrangement of the
anchors decides sufficiency for generic anchor values. Section 3.7
states a rank condition that separates a stable slope from a stable
slope and intercept across contexts.

All mathematical claims are conditional on the stated assumptions. Much
of the machinery is borrowed: affine linking, calibration, graph
synchronization, two-tree combinatorics and network least-squares
linking of many scales are not new (Section 6.2). The contribution is a
typed framework, made of gates that decide which overlaps may enter that
machinery and an account of what the resulting graph identifies. The
source of each borrowed mathematical step is credited in Appendix A.

The identification results and computational checks use no new
participant study. Section 4 applies the certification procedure to one
published participant-level dataset, comparing depiction styles within a
single method. Certifying a new empirical cross-method edge would need
estimates from both methods on shared stimuli and a joint uncertainty
model; paired participant-level records are one route to that.

Readers can take different routes through the paper. Vision scientists
can start with Section 2, the gates in Section 3.1 and Table 2, and the
worked example in Section 4. Mathematicians can read Section 3 with the
proofs in Appendix A. Readers from art history can read the terms and
the running example in Section 2, then Section 4 and the scope statement
in Section 6.1.

\hypertarget{measurement-operations-terms-and-the-running-example}{%
\section{2 Measurement operations, terms and the running
example}\label{measurement-operations-terms-and-the-running-example}}

Measurement transport, as used here, is a justified conversion of
coordinates or units from one measurement block to another; it is
distinct from causal transportability. Correlation alone cannot justify
it. One measurement can calibrate another only through admissible
transformation groups (the rescalings each scale permits; Stevens, 1946,
Table 1, p.~678, classifies scale types by these groups), shared
contrasts, nuisance fields and error models. In the relative-size study
(Wagemans, van Doorn, and Koenderink, 2011, ``Pictorial depth probed
through relative sizes''), size-adjustment and pointing configurations
correlated strongly for the observers studied, yet each retained an
observer-specific scale. The rank study (van Doorn, Koenderink, and
Wagemans, 2011, ``Rank order scaling of pictorial depth'') likewise
found close agreement in order without a corresponding relation between
ordinal resolution and pointing-derived range.

In pictorial relief, Koenderink, van Doorn, Kappers, and Todd (2001) had
four observers measure four photographed sculptures with four tasks,
from setting an attitude probe (the gauge figure; see the terms below)
to reproducing cross sections, and converted every result to a depth
map. Raw maps often disagreed across tasks. The authors then regressed
one map on another together with the image coordinates,
\(z_2=a+bx+cy+dz_1\). This ``affine regression'' removes an offset, a
depth shear and a depth gain. After it, the coefficients of
determination were generally between 0.8 and 1.0 for all tasks and
observers (pp.~442-445). The nuisance basis \([\mathbf 1,x,y]\) with a
positive block gain used below is that regression, so the empirical
finding that different depth tasks agree up to this ambiguity is theirs.
Their comparisons were pairwise fits judged by \(R^2\), which the added
terms can only raise. No margin was fixed in advance for how close
agreement must be, and no more than two maps were linked at once.

\textbf{Table 1. Operation type, admissible transformation, and
strongest claim before cross-method linking.} The entries are modeling
declarations for the operations as implemented in the sources examined
here, not theorems about those operations in every implementation.

\begin{longtable}[]{@{}
  >{\raggedright\arraybackslash}p{(\columnwidth - 6\tabcolsep) * \real{0.25}}
  >{\raggedright\arraybackslash}p{(\columnwidth - 6\tabcolsep) * \real{0.25}}
  >{\raggedright\arraybackslash}p{(\columnwidth - 6\tabcolsep) * \real{0.25}}
  >{\raggedright\arraybackslash}p{(\columnwidth - 6\tabcolsep) * \real{0.25}}@{}}
\toprule
Observed operation or output & Default admissible transformation &
Nuisance and context variables & Strongest claim without a cross-method
link \\
\midrule
\endhead
Pairwise depth order & Increasing transformations; parametric scale
confounded with discrimination & Order bias, lapse, observer, session,
comparison graph & Order and uncertainty; score differences only with
retained trial probabilities and a known link and discrimination
(Sections 3.6, 5.1) \\
Pointing-derived configuration & Offset, declared picture-plane field,
and gain as supported by the task & Observer, artwork, viewing geometry,
residualization basis & Gauge-reduced within-method coordinate
differences \\
Relative-size configuration from log size ratios & Positive affine
transformation & Observer/session scaling, selected locations, viewing
condition & Within-operation relative depth differences \\
Object volume or 3-D appearance rating & Response-model dependent &
Scale use, object set, medium, viewing condition & Construct-specific
rating effects \\
Foreground-background separation & Operation-specific ordinal or
interval model & Task instructions, segmentation, observer & Separation
under the same operation \\
Patch or material three-dimensionality rating & Rating-model dependent &
Patch selection, material, medium, response scale & Local
depicted-material response \\
Geometrical presence or viewing layout & Coordinate model conditional on
viewing geometry & Display, eye, stereo, frame, scene/picture relation &
Matched-condition layout \\
\bottomrule
\end{longtable}

The identification analysis uses the measurement block as its node. A
node key must include outcome type, method, observer or aggregation
level, artwork, depiction style, viewing condition, and processing
pipeline. A depiction style is any declared category of how a picture is
rendered, such as medium, technique or school; the framework treats it
as one field of the node key. Viewing condition is part of the node key:
blocks measured under different viewing conditions are different nodes,
and Table 1's context column lists what must be recorded to tell them
apart. The transformation groups in the table are modeling defaults to
be justified for each operation; they are not permissions to remove any
observed difference after seeing the result.

\textbf{Terms held to one meaning.} The following words keep one meaning
throughout.

\begin{itemize}
\tightlist
\item
  \emph{Method}: the observer's task, such as a relative-depth judgment,
  an attitude-probe setting, a relative-size adjustment or a rating. A
  statistical estimator or test is called a procedure, never a method.
\item
  \emph{Depiction style} (or \emph{style}): the declared rendering
  category of the stimulus, one field of the node key; in the running
  example, the six rendering conditions of Cole et al.~(2009).
\item
  \emph{Attitude probe}: the task in which a participant rotates a small
  disc at an image point until it appears to lie on the surface. Cole et
  al.~(2009) and Koenderink's group call it the gauge figure; this paper
  uses the word gauge only in the mathematical sense below.
\item
  \emph{Gauge}: a transformation of the parameters that leaves every
  observation unchanged, such as a block's own origin and unit or a
  declared nuisance term. Gauge reduction removes the declared nuisance
  terms (Section 3.1); a gauge constraint fixes the remaining free scale
  by a normalization (Section 3.3).
\item
  \emph{Chart}: a measurement block regarded as a coordinate system with
  its own unit and origin (Section 3.6). Counts of charts, such as
  \(M\), count blocks, not methods.
\end{itemize}

\textbf{Running example.} Cole et al.~(2009) deposited participant-level
attitude-probe settings for 12 rendered 3-D models, each shown in up to
six depiction styles: a shaded image, an artist's line drawing made from
it, occluding contours, apparent ridges, ridges and valleys, and
suggestive contours. In an attitude-probe task, a participant rotates a
small disc at a fixed image point until it appears to lie on the
surface. A model's probe points are identical across its styles, and
each participant saw only one style of any model. Within one method, the
data therefore give blocks scored on shared observations by independent
participants. The example concerns depiction-style blocks only, not art
in general or other methods.

\textbf{Nodes and edges.} A node is one measurement block: settings
averaged per probe point over the participants who saw one style of one
model, 70 blocks in all. A candidate edge joins two styles of one model
on the 85 to 178 probe points they share. Models share no probe point,
so these data can never place two models' blocks in one unit. Section 4
applies the six gates to these blocks.

\hypertarget{active-overlaps-and-a-common-relative-unit}{%
\section{3 Active overlaps and a common relative
unit}\label{active-overlaps-and-a-common-relative-unit}}

This section states the results; Appendix A holds the proofs and the
sources of each borrowed step. In the first four subsections, blocks
share a relative unit once a connected graph of active edges joins them.
The last three give the conditions for blocks to share one coordinate as
well, origin included: in the pairwise-anchor design and for generic
anchor values, two edge-disjoint spanning trees of anchors.

\hypertarget{model}{%
\subsection{3.1 Model}\label{model}}

Let \(V=\{1,\ldots,n\}\) index quantitative measurement blocks. For
candidate overlap \(e=(u,v)\), let \(d_e\in\mathbb{R}^{m_e}\) be a
shared latent configuration and let \(Q_e\) contain the declared
nuisance fields. The population mean in block \(v\) is

\[
\mu_{ve}=Q_e\gamma_{ve}+a_vd_e,\qquad a_v>0.
\]

Each block mean therefore splits into a prespecified nuisance component
plus the same edge-specific latent pattern multiplied by a positive
block gain. Both endpoint blocks are assumed to respond to the same
underlying residual pattern on the shared observations, differing only
by their nuisance terms and that gain.

For simple contrasts \(Q_e\) may contain only a constant. For pictorial
configurations compared across observers, a specified low-dimensional
approximation can use \(Q_e=[\mathbf 1,x,y]\).

Koenderink and van Doorn (2012) studied the global depth-gauge form
(unrelated to the attitude probe of Section 2): the depth
transformations by which different observers' pictorial spaces typically
differ (p.~1221). Including \([\mathbf 1,x,y]\) in the nuisance basis
removes an additive image-position component of that form. The basis
does not represent the full gauge action, which also contains
multiplicative depth scaling and admits space-varying generalizations.
Koenderink, van Doorn, Kappers, and Todd (2001) used the same three
terms with a depth gain to compare depth maps across tasks. The nuisance
basis must be selected independently of the scale comparison and applied
at both endpoints.

Let \(P_e\) be a common residualizing operator, meaning one that
annihilates the declared nuisance columns at both endpoints,
\(P_eQ_e=0\). Then

\[
\mu_{ve}^{\perp}=P_e\mu_{ve}=a_vd_e^{\perp},
\qquad
d_e^{\perp}=P_ed_e.
\]

Applying the same residualizer at both endpoints removes the declared
nuisance basis, so under the model the remaining endpoint profiles
differ only by their node gains. We call this removal gauge reduction;
``after the nuisance quotient'' below means after it.

The model also assumes that each node has one gain \(a_v\) across all
incident overlaps. If the gain changes with artwork, style, or
condition, those observations belong to separate nodes until a linking
model, a declared and tested relation between the two blocks' gains, is
accepted.

A gain belongs to the measurement block, not to the picture: it measures
how strongly that method, observer and condition express a shared depth
pattern. An edge needs observations both blocks share, so a gain ratio
compares how much depth two pictures elicit only where they share probed
points, as two depiction styles of one 3-D model do in Section 4.

An overlap is active only if it passes the six gates of Figure 1. The
outputs must be type-compatible (outcome type, gate 1). The declared
nuisance basis must be residualized at both endpoints (nuisance model,
gate 2). The two blocks must share a depth orientation (depth
orientation, gate 3). The shared residual pattern must be nonzero,
\(d_e^{\perp}\neq0\) (nonzero overlap, gate 4). Proportionality must be
sufficiently precise (proportionality, gate 5). Finally, the covariance
of the residualized paired estimates must be estimable well enough to
give the edge's log-ratio adequate information (covariance, gate 6). The
covariance gate is also a weak-identification check: the edge's
log-ratio must carry at least a prespecified minimum of information. A
shared depth orientation means that both blocks' scores run the same way
on the shared observations: a larger value means nearer in both, or
farther in both. Blocks scored in opposite directions would need a
negative gain ratio, which the positive gains \(a_v>0\) of the model
exclude; they fall outside the positive-gain model, and Section 3.4's
signed-gain remark names that case. ``Sufficiently precise'' does not
mean a nonsignificant lack-of-fit test. It means that an upper
confidence bound for a prespecified, normalized nonproportionality
measure falls below its prespecified equivalence margin. This check is
falsifiable only with at least two nonredundant residual contrasts
(independent directions of the shared profile left after nuisance
removal; with \(Q_e=[\mathbf 1,x,y]\), at least five shared
observations), or with replication that supplies independent contrasts.
An edge with an estimable ratio but inconclusive compatibility remains a
candidate rather than an active edge.

Table 2 lists each requirement with its operational check and the
consequence of failing it; Section 3.4 defines the statistic that gate 5
bounds.

\textbf{Table 2. Edge requirements, their operational checks, and the
consequence of failing each.}

\begin{longtable}[]{@{}
  >{\raggedright\arraybackslash}p{(\columnwidth - 4\tabcolsep) * \real{0.33}}
  >{\raggedright\arraybackslash}p{(\columnwidth - 4\tabcolsep) * \real{0.33}}
  >{\raggedright\arraybackslash}p{(\columnwidth - 4\tabcolsep) * \real{0.33}}@{}}
\toprule
Edge requirement & Operational check & If unmet \\
\midrule
\endhead
Outcome type (gate 1) and depth orientation (gate 3) & Match the node
keys and response operations before fitting a ratio; in data, a lower
confidence bound on the shared cross-product above zero (Section 4) &
Keep separate nodes or require an explicit linking model \\
Nuisance model (gate 2): declared common nuisance gauge & Specify
\(Q_e\) in advance and inspect residual structure & Do not interpret the
fitted gain as a transportable unit \\
Node-wide gain (precondition of the model, not one of the six gates;
repeated overlaps can contradict it) and nonzero residual signal (gate
4) & Check repeated overlaps and residual signal relative to error &
Split the node or delete an exact-zero edge; flag a weak edge \\
Proportionality within a margin (gate 5) & Use independent residual
contrasts and an upper confidence bound on normalized lack of fit &
Leave the edge as a candidate, even with a large lack-of-fit
\(p\)-value \\
Estimable joint uncertainty and a minimum of information (gate 6) & Use
paired records, a published covariance, or a justified joint model & Do
not certify an empirical edge from marginal errors alone \\
\bottomrule
\end{longtable}

\hypertarget{gauge-reduced-active-overlap-theorem}{%
\subsection{3.2 Gauge-reduced active-overlap
theorem}\label{gauge-reduced-active-overlap-theorem}}

In plain terms, each active edge fixes the ratio of two blocks' gains,
so a chain of active edges fixes every gain relative to one reference,
and blocks in different connected pieces of the graph stay on unrelated
scales. Figure 2 illustrates the setting on synthetic profiles: removing
the declared nuisance terms leaves one shared residual shape (panel a),
and a graph with two components keeps two unrelated scales (panel b).

Let \(G^*=(V,E^*)\) be the active-overlap graph, whose nodes are the
measurement blocks and whose edges are the active overlaps, and
\(c(G^*)\) its number of connected components.

\textbf{Theorem 1.} Assume the model of Section 3.1, so that the
residual endpoint profiles of each active edge are exactly proportional;
assume also that the nuisance gauge is declared, common residualization
is valid at both endpoints of each edge, each node has a gain shared
across its incident overlaps, gains are positive, and
\(d_e^{\perp}\neq0\) on every active edge. Each active edge \(e=(u,v)\)
then contributes one edge equation \(\alpha_v-\alpha_u=r_e\), where
\(\alpha_v=\log a_v\) is node \(v\)'s log gain and \(r_e=\log\rho_e\) is
the log of the edge's positive proportionality ratio \(\rho_e\). Then:
the Jacobian of the edge equations has rank \(n-c(G^*)\); the remaining
multiplicative scale freedom, the residual scale fiber, has dimension
\(c(G^*)\), counted after the declared nuisance quotient (the raw
\(\gamma_{ve}\) and \(d_e\) keep further additive directions,
\(d_e\to d_e+Q_et\) with \(\gamma_{ve}\to\gamma_{ve}-a_vt\), which the
quotient removes); one normalization per component identifies the
remaining relative gains; and a connected active-overlap graph places
all node gains in one common relative unit after the declared nuisance
quotient. The relative-gain statements need only the assumptions in the
theorem's first sentence, which are sufficient for them.

Theorem 1 itself is not new. Once logarithms turn the gain ratios into
differences, it is the identifiability statement for synchronization on
a graph (Singer, 2011); Appendix A.1 gives the proof and that
correspondence.

The specific contribution is therefore the graph, not the theorem.
Singer's measurement graph arrives already built. Ours does not. An
overlap becomes an active edge only after it passes the six gates of
Figure 1 and Section 3.1. The gains the theorem links belong to
residualized endpoint profiles, after the declared nuisance quotient,
rather than to raw measurements. The contribution consists of the rules
for admitting overlaps as edges and the interpretation of nodes;
applying the synchronization result to that graph then identifies
relative node gains.

\hypertarget{network-estimation-and-cycle-diagnostics}{%
\subsection{3.3 Network estimation and cycle
diagnostics}\label{network-estimation-and-cycle-diagnostics}}

Let \(r\) collect estimated log edge ratios, let \(B\) be the oriented
node-by-edge incidence matrix of \(G^*\). Let \(\Sigma_r\) be the
covariance of \(r\). With node log gains \(\alpha\) and a gauge
constraint (a normalization fixing the one free scale in each component)
\(C\alpha=0\), estimate the linked units by

\[
\widehat{\alpha}
=
\arg\min_{C\alpha=0}
(r-B^\top\alpha)^\top
\Sigma_r^{-1}
(r-B^\top\alpha).
\]

The estimator chooses node log-gains whose predicted edge differences
best match the observed log-ratios, weighting each discrepancy by the
joint covariance of the estimates. Its unweighted relative is the
least-squares linking of many separately calibrated test forms in
Haberman (2009); Battauz (2013, 2017) gives multi-path and simultaneous
versions with standard errors.

If edge errors are independent with \(\Sigma_r=W^{-1}\), the information
matrix is the weighted graph Laplacian \(L_W=BWB^\top\).

Connectivity determines estimability, while bridge weights determine the
uncertainty of named contrasts such as \(\alpha_v-\alpha_u\). A formally
connected graph can therefore contain a link too weak to support a
useful comparison. Proposition 1 makes that distinction exact.

\textbf{Proposition 1 (precision of an estimable contrast).} Let
\(J=B\Sigma_r^{-1}B^\top\) be the information operator of the network
estimator. For any contrast \(c\) that is estimable, meaning \(c\) is
orthogonal to the vectors constant within each component,

\[
\operatorname{Var}(c^\top\widehat\alpha)=c^\top J^{+}c,
\]

with \(J^{+}\) the Moore-Penrose pseudoinverse. Writing \(\iota_v\) for
the indicator vector of node \(v\), if the graph is connected and edge
errors are independent, so that \(\Sigma_r=W^{-1}\) and \(J=L_W\), then
for a node pair

\[
\operatorname{Var}(\widehat\alpha_v-\widehat\alpha_u)
=(\iota_v-\iota_u)^\top L_W^{+}(\iota_v-\iota_u),
\]

the weighted effective resistance between \(u\) and \(v\) in the network
whose conductances are the edge information weights.

In words: connectivity is the yes-or-no identification condition, and
effective resistance is the target-specific precision condition. A
connected graph can identify \(\alpha_v-\alpha_u\) and still leave it
nearly useless, because a long chain of low-weight edges between \(u\)
and \(v\) has high resistance. Effective resistance is the exact content
of the paper's ``weak link'' language. Appendix A.2 gives the proof and
the sources of this identity.

A design consequence follows immediately. The independence assumption
below is sufficient for the displayed formula, not necessary for a new
edge to add information. The formula does not cover a new edge whose
error is correlated with existing edges. Suppose a candidate new edge
joins blocks \(i\) and \(j\) that already lie in the same connected
component, with incidence vector \(g=\iota_j-\iota_i\), information
weight \(w>0\) and an error independent of the existing edges, so that
the Laplacian becomes \(L+wgg^\top\). Then for any estimable target
contrast \(c\) the variance falls by exactly

\[
\Delta V(c)
=
\frac{(c^\top L^{+}g)^2}
{w^{-1}+g^\top L^{+}g}\ \ge\ 0,
\]

by the Sherman-Morrison identity applied on the estimable subspace. The
condition that \(i\) and \(j\) already share a component is necessary:
it is what keeps \(g\) inside that subspace. An edge between two
separate components does something else. It makes a new relative gain
estimable and leaves the variance of every previously estimable contrast
unchanged, so the formula, which would report a positive reduction for
such an edge, does not apply to it.

The denominator adds the candidate edge's own measurement noise to the
effective resistance already between its endpoints, and the numerator
measures how heavily the target contrast draws on the potential
difference across that edge. Here the expression scores a prespecified
cross-method calibration contrast. A design can check two gates before
collecting data: outcome type and nuisance model (gates 1 and 2). Among
candidate edges that pass those two, this quantity ranks them for a
prespecified target, given an assumed information weight \(w\). The
ranking is conditional: an edge delivers the reduction only if its data
later pass gates 3 to 6, and a mathematically valuable edge is not
thereby a scientifically valid one. Appendix A.3 says how this score
differs from the edge derivatives of Ghosh, Boyd, and Saberi (2008).

Let \(\widehat r=B^\top\widehat\alpha\) be the fitted edge field and
\(\tilde r=r-\widehat r\) the residual field. The statistic

\[
Q_{\rm cycle}=\tilde r^\top\Sigma_r^{-1}\tilde r
\]

measures global disagreement with the node-gain model. Its geometry is
explicit: the estimator's normal equations give
\(B\Sigma_r^{-1}\tilde r=0\), so the fitted edge field \(\widehat r\),
which lies in the image of \(B^\top\), and the residual field
\(\tilde r\) are orthogonal in the covariance-weighted inner product.
The fit is therefore the covariance-weighted projection of the observed
edge log-ratios onto the space of fields that some assignment of node
gains could have produced, and \(Q_{\rm cycle}\) is the squared weighted
length of the part of the observed field that no such assignment can
produce.

The link between this projection and the weighted graph-Hodge
decomposition of Jiang, Lim, Yao, and Ye (2011), and the conditions
under which the graph reading survives correlated edge errors, are in
Appendix A.4. \(Q_{\rm cycle}\) is a covariance-aware cycle-closure
diagnostic: it can test overidentifying restrictions when cycles are
present, but it cannot validate construct compatibility or rescue a
misspecified nuisance quotient. Under a correctly specified Gaussian
model with known nonsingular covariance and a fixed active-edge set, its
reference degrees of freedom are \(|E^*|-n+c(G^*)\), the dimension of
the active-overlap graph's cycle space. If the same observations select
or certify edges and then enter \(Q_{\rm cycle}\), this reference
distribution ignores selection. Valid calibration then requires sample
splitting, selective inference, or a resampling procedure that repeats
the entire edge-selection and testing pipeline. The diagnostic is
informative only when each edge log-ratio is estimated from that edge's
own overlap data. Ratios computed as differences of one set of fitted
node gains close every cycle by construction; their covariance is
singular on the cycle space, and they give \(Q_{\rm cycle}=0\) whatever
the data. The omnibus statistic needs no cycle-by-cycle multiplicity
correction; separately reported edge or cycle tests do.

\hypertarget{corollaries-and-the-gate-5-proportionality-statistic}{%
\subsection{3.4 Corollaries and the gate 5 proportionality
statistic}\label{corollaries-and-the-gate-5-proportionality-statistic}}

\textbf{Cycle closure:} population log ratios sum to zero around every
cycle; estimated ratios need an uncertainty-aware diagnostic, not exact
equality.

\textbf{Zero bridge:} if \(d_e^{\perp}=0\), the residual profiles carry
no information about the gains (their derivative with respect to a gain
is \(d_e^{\perp}\), which is zero), and the edge is removed by the
nonzero-overlap gate (gate 4).

\textbf{Weak bridge:} information weights approach zero with residual
signal-to-noise; formal rank can coexist with unusable uncertainty.

\textbf{One-contrast distinction:} one nonzero scalar contrast can
identify a ratio under the model but cannot test proportionality.

\textbf{Signed gains:} allowing negative gains replaces the positive
log-gain model with a signed-graph problem.

A reference structure for a preregistered certification procedure is as
follows; an application must also specify its resampling unit,
covariance estimator, equivalence margin, and the minimum of information
that the covariance gate (gate 6) requires. Prespecify \(Q_e\),
residualize both endpoints, and fit the positive proportionality
coefficient \(\rho_e\). With \(y_u\) and \(y_v\) the residualized
observed endpoint profiles and a declared weight matrix \(W_e\) on the
\(m_e\) profile coordinates (not the edge weights \(W\) of Section 3.3),
define

\[
\Delta_e
=
\frac{\|y_v-\rho_ey_u\|_{W_e}}
{\|y_v\|_{W_e}+\|\rho_ey_u\|_{W_e}}.
\]

Evaluate \(\Delta_e\) only when its denominator is positive. With a
positive-definite \(W_e\), the denominator vanishes only when both
observed profiles are zero; there \(\Delta_e\) is undefined, not zero,
and the edge stays uncertified. Whenever \(\Delta_e\) is defined, the
triangle inequality gives \(0\le\Delta_e\le1\). Thus \(\Delta_e=0\)
means exact proportionality, and larger values measure residual shape
disagreement relative to the overall fitted signal magnitude. Obtain a
profile or bootstrap upper confidence bound \(U_e\). Certify
proportionality only if \(U_e<\delta_e\) for a prespecified equivalence
margin \(\delta_e\), and independently require a prespecified lower
bound on the information for the edge's log-ratio \(r_e\) (the
covariance gate, gate 6). The proportionality gate (gate 5) and the
covariance gate answer different questions, and both must pass. Passing
them is evidence that the two profiles are proportional to within
\(\delta_e\), not proof of exact proportionality, because
\(\delta_e>0\). The gain identification of Theorem 1 applies to a
certified edge under the model assumption that the population profiles
are exactly proportional, and the fitted gains carry that assumption.

\includegraphics{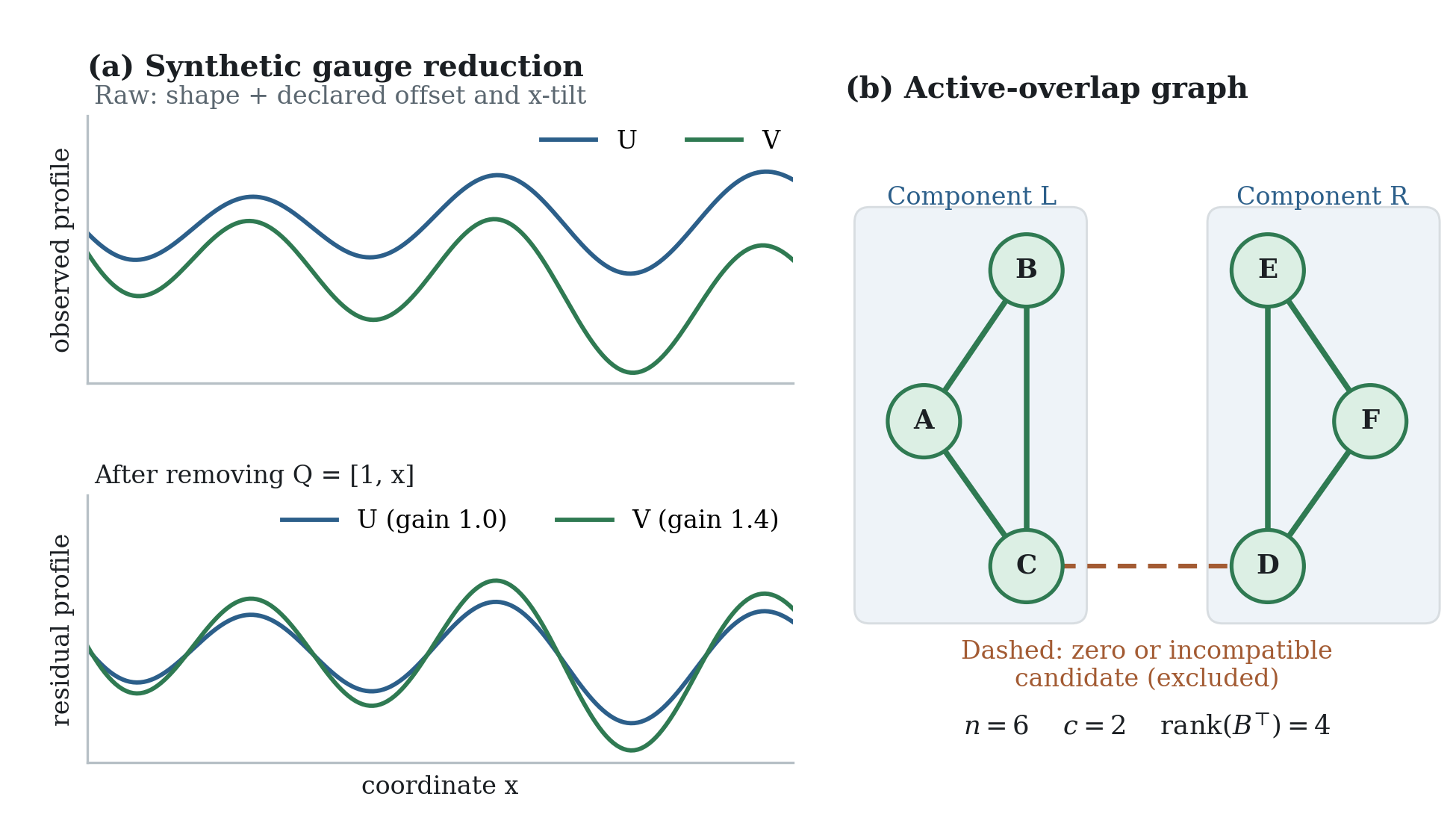}

Figure 2. Gauge reduction and graph connectivity. Synthetic profiles
illustrate how removing a declared offset and one-coordinate tilt leaves
a common residual shape with different gains. The graph panel
illustrates the rank result: the dashed zero or incompatible candidate
does not connect the two components. The illustration uses
\([\mathbf 1,x]\), whereas the simulation in Appendix B uses
\([\mathbf 1,x,y]\); neither establishes that an affine gauge is
adequate for every picture.

\hypertarget{two-identification-problems-that-must-not-be-conflated}{%
\subsection{3.5 Two identification problems that must not be
conflated}\label{two-identification-problems-that-must-not-be-conflated}}

\textbf{Table 3. Transported objects and limits of the two
identification results.}

\begin{longtable}[]{@{}
  >{\raggedright\arraybackslash}p{(\columnwidth - 4\tabcolsep) * \real{0.33}}
  >{\raggedright\arraybackslash}p{(\columnwidth - 4\tabcolsep) * \real{0.33}}
  >{\raggedright\arraybackslash}p{(\columnwidth - 4\tabcolsep) * \real{0.33}}@{}}
\toprule
Question & Active-overlap theorem (Theorem 1) & Rank criterion (Lemma
1) \\
\midrule
\endhead
Shared object & Edge-specific residual configurations & Common stimulus
scores \(z_i\) \\
Transported quantity & Node gain or relative unit & Affine coordinate
system and common scalar representation \\
Rank object & Graph incidence matrix & Full chart-incidence system \\
Remaining freedom & One gain normalization per component & One global
positive affine transformation \\
What it does not establish & A universal stimulus coordinate & Adequacy
of the affine model in noisy data \\
\bottomrule
\end{longtable}

Table 3 sets the two results side by side. Theorem 1 links relative
units. The rank criterion asks the stronger question of whether all
in-scope charts can be placed on one common coordinate. In both cases,
compatibility or model existence comes before rank, and precision comes
after identification.

\hypertarget{mixed-method-identification}{%
\subsection{3.6 Mixed-method
identification}\label{mixed-method-identification}}

A complementary affine incidence model makes the shared-stimulus
assumptions explicit and handles mixed methods. Here affine chart means
an affine coordinate parameterization, not a differential-geometric
chart. A chart is one measurement block of Section 2 with its own unit
and origin, so two observers using the same method are two charts unless
they are declared to share one calibration. Every count in this section,
\(M\) included, counts charts (measurement blocks), not methods.

Let \(I\) index \(N\) in-scope stimuli and \(\mathcal M\) index \(M\)
charts. For every observed pair \((m,i)\), write

\[
z_i=\lambda_mx_{mi}+\nu_m,
\qquad
\lambda_m>0.
\]

Here \(\lambda_m\) is a chart slope, the reciprocal of a gain, and
\(\nu_m\) a chart intercept. Every chart-specific coordinate is
therefore allowed its own positive change of unit and its own change of
origin, but all charts must map their shared stimuli to the same latent
scalar \(z_i\).

The model has no image-position terms. Depth maps that carry
observer-specific depth shear (Sections 2 and 3.1) satisfy it only after
the declared nuisance basis has been removed identically at every chart;
the coordinate identified is then the residual coordinate, not the raw
depth map.

Stacking these equations gives \(K\eta=0\), where

\[
\eta=(z_1,\ldots,z_N,\lambda_1,\nu_1,\ldots,\lambda_M,\nu_M).
\]

\textbf{Lemma 1 (rank criterion).} Assume every in-scope stimulus is
scored by at least one chart and that an exact compatible solution with
positive slopes exists. The common representation and all chart
calibrations are identified up to one global positive affine
transformation exactly when \(\dim\ker K=2\), equivalently
\(\operatorname{rank}K=N+2M-2\). The two permitted null directions are
exactly the unavoidable global change of origin and positive change of
unit; any third null direction is an additional nonidentifiability.

For each stimulus, eliminate its \(z_i\) using one incident chart and
compare the remaining charts to that reference. If \(H\) stacks these
overlap equalities, row elimination gives

\[
\operatorname{rank}K=N+\operatorname{rank}H.
\]

Thus a stimulus read by all \(M\) charts contributes at most \(M-1\)
independent raw equalities. Call a stimulus scored by two or more charts
an anchor: it ties those charts together, and \(z_e\) denotes anchor
\(e\)'s common latent score. Counting all \(\binom{M}{2}\) pairwise
restatements as separate anchors would overstate the available rank.

If each anchor is scored by exactly two charts and different edges use
different stimuli, fix one exact solution \((\lambda^0,\nu^0)\) and
change calibration coordinates to

\[
\tilde\lambda_m=\lambda_m/\lambda_m^0,
\qquad
\tilde\nu_m=\nu_m-\nu_m^0\tilde\lambda_m.
\]

In these coordinates the row for an anchor \(e\) shared by charts \(m\)
and \(n\) reads
\(z_e(\tilde\lambda_m-\tilde\lambda_n)+(\tilde\nu_m-\tilde\nu_n)=0\), so
this invertible change puts \(H\) in the form

\[
H(z)=[D(z)A\mid A],
\qquad
D(z)=\operatorname{diag}(z_e),
\]

with \(A\) the oriented edge-by-node incidence matrix of the anchor
graph, whose nodes are the \(M\) charts and which has one edge for each
anchor stimulus, so that several anchors shared by the same two charts
are parallel edges. \(A\) is built the same way as \(B^\top\) in Section
3.3 but on a different graph: Section 3.3's edges are certified overlaps
between blocks, each carrying a whole residual profile. An anchor edge
is one scalar equation, not a gated overlap: a single shared stimulus
gives each block one number, so proportionality cannot be tested on it,
and the results below assume compatibility (Lemma 1) rather than certify
it. Each anchor row therefore contains two copies of the incidence
pattern, one weighted by that anchor's coordinate and one unweighted,
constraining both the slope-like and the intercept-like calibration
coordinates of the two charts the anchor joins. If the \(z_e\) can vary
independently, a full-size minor is a nonzero polynomial precisely when
the graph contains two edge-disjoint spanning trees. This existence
condition is the two-tree case of the Nash-Williams--Tutte spanning-tree
packing theorem. The direct determinant proof, and the link between this
matrix pattern and the general frames of Whiteley (1988), are in
Appendix A.5.

The determinant expansion of Appendix A.5 says more than whether the
design has full rank. For algebraically generic anchor values it says
exactly how far short a deficient design falls, which is the quantity a
designer actually needs; a realized anchor assignment must still be
checked separately.

\textbf{Proposition 2 (generic rank and partition deficit).} In the
pairwise-anchor design, for algebraically generic anchor values, a set
of edge rows of \(H(z)\) is linearly independent exactly when it can be
partitioned into two forests of the anchor graph. The generic row
structure is therefore the union of two copies of that graph's cycle
matroid, and the generic rank is

\[
r_{\rm gen}
=
\min_{F\subseteq E}
\bigl\{\,|E\setminus F|+2\bigl(M-c(F)\bigr)\,\bigr\},
\]

where \(c(F)\) counts the connected components of the spanning subgraph
\((V,F)\), isolated charts included. Equivalently, writing
\(\operatorname{cr}(\mathcal P)\) for the number of anchor edges joining
different parts of a partition \(\mathcal P\) of the \(M\) charts, the
number of calibration directions left unidentified beyond the global
affine gauge (the one global change of origin and unit that no data can
fix) is

\[
D_{\rm gen}
=
2(M-1)-r_{\rm gen}
=
\max_{\mathcal P}
\bigl\{\,2(|\mathcal P|-1)-\operatorname{cr}(\mathcal P)\,\bigr\}.
\]

The result is a counting rule. Split the charts into any number of
groups. Merging those groups into one calibrated system needs two
crossing anchors for each merge, that is \(2(|\mathcal P|-1)\) of them,
and, for algebraically generic anchor values, the largest shortfall over
all ways of splitting is exactly how many affine directions remain
unidentified beyond the global affine gauge. The three designs computed
in Appendix B.4 are the formula's first three cases: the three-edge tree
on four charts has singleton-partition deficit \(2(4-1)-3=3\) and hence
generic rank \(6-3=3\), the full \(2(M-1)=6\) less the deficit; the
two-component doubled-edge design has deficit \(2\); the well-spread
\(K_4\) has deficit \(0\). Full generic identification is the
zero-deficit case, which by the Nash-Williams--Tutte packing theorem is
the two-edge-disjoint-spanning-trees condition already stated. Appendix
A.6 gives the proof and separates the parts of Proposition 2 that are
Whiteley's (1988) or standard matroid-union theory from the deficit
reading this paper adds.

A specified assignment can still be exceptional. For \(K_4\), set
\(z_{ij}=u_i+u_j\) with \(u=(0,1,3,7)\). All six anchors are distinct,
yet the perturbation

\[
\Delta\tilde\lambda_i=u_i,
\qquad
\Delta\tilde\nu_i=-u_i^2
\]

in the coordinates above satisfies every edge equation, adding one
nongauge null direction. Exact symbolic checks give calibration rank
five for this assignment, so beyond the two unavoidable affine gauge
directions one further direction is unidentified, leaving nullity three.
A checked nonexceptional integer assignment instead has a selected
gauge-fixed full minor with \(|\det|=40\), nonzero and therefore
certifying full calibration rank for that realized assignment. The
additive construction is sufficient to produce an extra nongauge null
direction, but it does not characterize every exceptional assignment:
the assignment \(z_e=1,2,4,8,16,32\) on edges \(01,02,03,12,13,23\) is
also exceptional (rank five), and the exceptional set is the full
determinant-zero variety. The next proposition shows that this
exceptional set has Lebesgue measure zero.

\textbf{Proposition 3 (exceptional anchor geometry is exceptional).} If
the anchor graph contains two edge-disjoint spanning trees, then at
least one gauge-fixed full-size minor of \(H(z)\) is a nonzero
polynomial in the anchor coordinates, by the construction in the proof
of Proposition 2. Rank deficiency requires every such minor to vanish,
so the rank-deficient anchor assignments lie in the zero set of that one
nonzero polynomial. They therefore form a proper algebraic subset of the
anchor space and have Lebesgue measure zero: under any anchor design
whose coordinate vector has a joint distribution with a density on the
anchor space, exact rank loss from anchor geometry has probability zero.
A density for each coordinate separately is not enough: the additive
assignment \(z_{ij}=u_i+u_j\) of Section 3.6 gives every coordinate a
density when \(u\) has one, yet always loses rank. \(\square\)

Proposition 3 must not be read as reassurance that conditioning is
almost always fine. It concerns exact rank only. An anchor assignment
can sit arbitrarily close to the exceptional set, and the smallest
nonzero singular value of the calibration matrix then approaches zero by
continuity, with all the instability that implies for finite samples.
Nor is a raw singular value itself a scientifically meaningful index:
changing the common affine coordinate or rescaling columns changes the
singular values of a chosen gauge-fixed matrix while leaving algebraic
identifiability untouched. A conditioning diagnostic is interpretable
only after a prespecified numerical gauge and column standardization, or
when it is computed from the Fisher information that the declared noise
model actually induces. Such a diagnostic is the honest bridge from
generic identification to weak identification, and it is a weaker bridge
than the probability-zero result alone would suggest.

The target \(2M-2\) is global and therefore presupposes a connected
anchor graph. If the graph has \(c\) components, independent affine
gauges remain in those components and the componentwise maximum
calibration rank is \(2(M-c)\). A disconnected design can attain that
componentwise maximum and still lack the \(2(c-1)\) dimensions required
to link its components into one global coordinate.

\textbf{Corollary 1: realized deficiency and compatibility
restrictions.} Let \(r=\operatorname{rank}H\) for a realized
pairwise-anchor design with \(M\) method-specific charts, \(E\) anchor
equations, and \(c\) anchor-graph components. The componentwise target
dimension is \(d_c=2(M-c)\), so the unidentified dimension within
components is \(u_{\rm comp}=d_c-r\). Relative to a single global affine
coordinate, the target is \(d_{\rm global}=2M-2\), and

\[
u_{\rm global}
=
d_{\rm global}-r
=
2(c-1)+u_{\rm comp}.
\]

The global deficiency therefore separates into two sources: two affine
degrees of freedom for every component beyond the first, plus whatever
additional rank is lost inside components.

The number of algebraic compatibility restrictions is \(f=E-r\). Thus
\(r=d_c\) gives componentwise identification; global identification
additionally requires \(c=1\). A connected design is just identified
when

\[
r=2M-2
\qquad\text{and}\qquad
f=0,
\]

and overidentified when

\[
r=2M-2
\qquad\text{and}\qquad
f>0.
\]

The counts \(r\) and \(f\) are algebraic. Statistical test degrees of
freedom equal \(f\) only under an appropriate fixed-design stochastic
model with the required covariance and regularity conditions. Exact
interpolation when \(f=0\) is not evidence that the affine model fits.

\textbf{Definition: deletion robustness.} A design is
leave-one-anchor-out identified if it has calibration rank \(2M-2\) and
deleting any single anchor row leaves calibration rank \(2M-2\). This is
stronger than minimal identification and provides a direct matrix
criterion for designs intended to survive one unusable or incompatible
bridge. Corollary 2 gives the generic graph version.

\textbf{Corollary 2 (leave-one-anchor-out criterion).} A generic
pairwise-anchor design stays globally identified after the deletion of
any single anchor edge exactly when every partition \(\mathcal P\) of
the charts into two or more parts satisfies

\[
\operatorname{cr}(\mathcal P)\ \ge\ 2(|\mathcal P|-1)+1.
\]

The design reading is a margin of one crossing anchor everywhere. Taking
the partition into singletons gives the necessary count \(E\ge2M-1\),
against \(E\ge2M-2\) for bare generic identification, so surviving the
loss of one anchor costs exactly one more anchor at minimum. The count
alone is not sufficient: every partition must carry the extra edge. A
simple \(K_4\) has \(E=6=2M-2\) and so can be just identified but can
never be robust to losing an anchor; a multigraph with additional
independent anchor observations on the same chart pairs can be. Appendix
A.7 gives the proof of Corollary 2.

Strict ordinal statements \(Cz>0\) define an open subset of the affine
solution space at an interior solution. They can orient or restrict the
solution, but its local unidentified dimension after quotienting the
two-dimensional affine gauge remains \(\dim\ker K-2\). A probabilistic
comparison method can supply more information: if
\(p_{ij}=F\!\left(d(z_i-z_j)\right)\), with known strictly increasing
invertible \(F\), positive discrimination \(d\) (as in Section 5.1), and
a connected comparison graph, the inverse-link edge values
\(F^{-1}(p_{ij})\) identify node scores \(u_i=dz_i+c\) up to one
additive constant. These are inverse-link values for a general \(F\),
not logits, which are specific to the logistic link. With known link
shape and discrimination, a comparison probability is therefore not
merely an order statement: after the inverse link is applied it supplies
a quantitative score difference. Those scores are an affine chart. The
result depends on this probability model and on retaining probabilities
or trials; a final ordering alone retains only inequalities.

\hypertarget{context-transport-has-slope-and-full-affine-levels}{%
\subsection{3.7 Context transport has slope and full-affine
levels}\label{context-transport-has-slope-and-full-affine-levels}}

A context is any declared condition under which the same methods are
run, such as a depiction style, viewing condition or observer group.
Section 5.2's styles are contexts, and its test is the statistical form
of the slope condition below.

For context \(s\), let \(\mu_{is}=b_s+a_sz_{is}\) be the \(M\)-vector of
method means, with positive gains \(a_s\) (entry \(m\) of \(a_s\) is the
forward slope \(1/\lambda_m\) of method \(m\)'s chart in context \(s\),
Section 3.6) and nonconstant \(z_{is}\) within each context. Within each
context the vector of method means therefore lies on an affine line with
direction \(a_s\), which is what makes the rank condition below natural.
Center each context's response columns and concatenate the resulting
matrices as \(L_{\rm ctx}\) (not a Laplacian). Then
\(\operatorname{rank}L_{\rm ctx}=1\) exactly when all \(a_s\) lie on one
positive ray, the condition for a shared slope conversion.

A shared full affine conversion additionally requires the context means
to have the same projection into \(\mathbb R^M/\operatorname{span}(a)\),
where \(a\) is the common slope direction; equivalently, all pooled
response columns lie on one affine line with a positive direction. In
plain terms, context-specific intercept vectors may differ only along
the common slope direction; any intercept shift orthogonal to that
direction prevents one full affine conversion from working across
contexts.

For two methods, write \(\mu_{2is}=r_s\mu_{1is}+c_s\). Slope transport
requires constant \(r_s\); full affine transport requires both constant
\(r_s\) and constant \(c_s\). The contexts \((\mu_1,\mu_2)=(z,2z)\) and
\((\mu_1,\mu_2)=(z,2z+5)\) satisfy the slope condition and fail the full
condition.

Both transport conditions concern conditional means. Meredith's
measurement-invariance framework also concerns conditional
distributions, including residual behavior, so the mean theorem does not
establish full distributional invariance.

\hypertarget{worked-example-depiction-styles-in-one-published-attitude-probe-dataset}{%
\section{4 Worked example: depiction styles in one published
attitude-probe
dataset}\label{worked-example-depiction-styles-in-one-published-attitude-probe-dataset}}

The relative-size study explicitly models depth configurations up to
positive affine transformation, estimates observer and session scaling
factors, and reports substantial heterogeneity in those factors. It also
distinguishes the absence of a natural origin from the absence of a
natural unit. These source-supported observations motivate separate
treatment of offset and gain; they do not certify a universal conversion
to another method.

The method-by-style study (van Doorn et al., 2012, ``Interaction of
depth probes and style of depiction'', p.~528) reports that, across the
sampled pictures, the relation between its two depth-measurement methods
depended on depiction style for all but one observer. A method relation
estimated in one style therefore should not be presumed to transport
unchanged to another. Applying the certification procedure to that study
would still require a joint uncertainty model for its two methods, for
which paired participant-level records are one route, and cross-style
covariance. The rest of this section applies it to the Cole et
al.~(2009) dataset of Section 2, a published dataset that includes
participant-level records, for depiction styles within one method.

\hypertarget{data-preparation-and-prespecified-rules}{%
\subsection{4.1 Data preparation and prespecified
rules}\label{data-preparation-and-prespecified-rules}}

\textbf{Gauge reduction (nuisance removal, Section 3.1).} Each setting
gives a surface normal, which we convert to a depth gradient. On a
gradient, the nuisance basis \([\mathbf 1,x,y]\) of Section 3.1 acts as
a constant offset and a depth gain as a multiplier. Differentiating a
depth map \(z(x,y)\) turns an added plane \(b_0+b_xx+b_yy\) into the
constant vector \((b_x,b_y)\), so the implemented residualizer subtracts
the mean of the \(p\) component and the mean of the \(q\) component
separately, rather than applying the depth-map matrix
\([\mathbf 1,x,y]\) directly. Together they form the bas-relief
ambiguity (Belhumeur, Kriegman, and Yuille, 1999, p.~33) that Cole et
al.~discuss (p.~2): from a shaded image alone, an observer cannot tell
whether a surface has been stretched or flattened in depth and sheared
(a plane added to its depth). The node gain \(a_v\) thus measures how
much relief the participants behind a block perceived in the shared
shape, relative to the blocks it is linked to. Residualizing removes the
offsets at both endpoints (gate 2); an offset added to one block changes
no residualized ratio (largest change \(3\times10^{-16}\)), against a
median change of 0.12 without residualizing.

\textbf{Prespecified rules.} Before fitting any ratio we fixed the
covariance gate at a standard error of at most 0.10 for each log ratio.
We fixed the proportionality margin at \(\delta=0.20\) for a
noise-corrected \(\Delta_e\), which is Section 3.4's statistic with each
squared norm reduced by its expected measurement-noise contribution and
the numerator floored at zero (each endpoint's corrected squared norm is
also floored at zero, and an undefined statistic counts as a gate 5
failure), so that two noisy copies of one shape score near 0. The
expected noise contribution of each profile entry is its variance of the
mean, the across-participant variance divided by the number of
participants, scaled by \((m-1)/m\) for the mean removed over the \(m\)
shared probe points. The ratio \(\rho_e\) is a Deming
(errors-in-variables) slope whose error-variance ratio is the \(v\)
block's average entry variance over the \(u\) block's. For this ratio,
the uncorrected statistic lies between \(\sin(\theta/2)\) and
\(\tan(\theta/2)\) when the two residual profiles, treated as vectors
over all shared probe points, meet at an acute angle \(\theta\);
Appendix A.8 derives this bracket. Before noise correction, the margin
therefore requires their directions to differ by less than about 23
degrees, an angle between whole profiles, not a difference in surface
slant. Because styles are between-subject, endpoint errors are
independent, and the covariance comes from 1,000 participant resamples
(seed 20261015), the same resamples for every edge of a model, so edges
that share a block keep their covariance. The noise correction treats
the fitted Deming ratio as fixed and takes the entries' noise as
independent, so the bracket above and the range bounds of Section 3.4
hold only before correction; the controls of Section 4.4 are the check
on the corrected statistic. The prespecification also fixed the
pipeline: probe points whose ground-truth slant is at most 75 degrees,
at least 5 participants per probe point in each block, settings capped
at 80 degrees of slant, and an orthographic conversion to depth
gradients that ignores the 30-degree field of view.

\hypertarget{results-under-the-prespecified-rules}{%
\subsection{4.2 Results under the prespecified
rules}\label{results-under-the-prespecified-rules}}

\textbf{Gate results.} Each gate was applied to every candidate edge, so
one edge can fail several. Of 171 candidate edges, all passed depth
orientation (gate 3) and nonzero overlap (gate 4). Proportionality (gate
5) passed for 2 and the covariance gate (gate 6) for 5. Each edge's gate
5 decision uses its own 95\% bound, not adjusted for testing 171 edges.
One edge passed all six prespecified screening rules (gate 1 holds by
design, since every block comes from one task): shaded image to ridges
and valleys on the cubehole model, with gain ratio 1.01, ridges and
valleys over shaded (log-ratio standard error 0.048; Figure 3C). This is
a rule-passing result, not a validation of gate 4's nominal error rate
(Section 4.4) and not a family-wise guarantee. The active-overlap graph
thus has 69 components on 70 nodes, and no cycle remains to test.

\textbf{Second implementation.} A second implementation re-derived every
gate decision. Claude (Anthropic), working in a separate session as the
Use of AI tools statement describes, wrote it in Python with NumPy from
the prespecification, the raw archive and the text of Sections 3.4 and
4, without opening the original code. The two implementations share no
code. Upstream they share the archive's convention for turning a setting
into a surface normal, the prespecified conversion of normals to depth
gradients, and the Python and NumPy versions, so an error in that shared
specification would reach both. Run with its own bootstrap seed, the
second implementation certified the same single edge, with the same gain
ratio (1.01) and a log-ratio standard error of 0.047 against the
original's 0.048. It agreed on all 171 edges for gates 3 to 5 and on 170
of 171 for gate 6, where the one difference is an edge whose standard
error lies at the 0.10 threshold.

\begin{figure}[tp]

\includegraphics{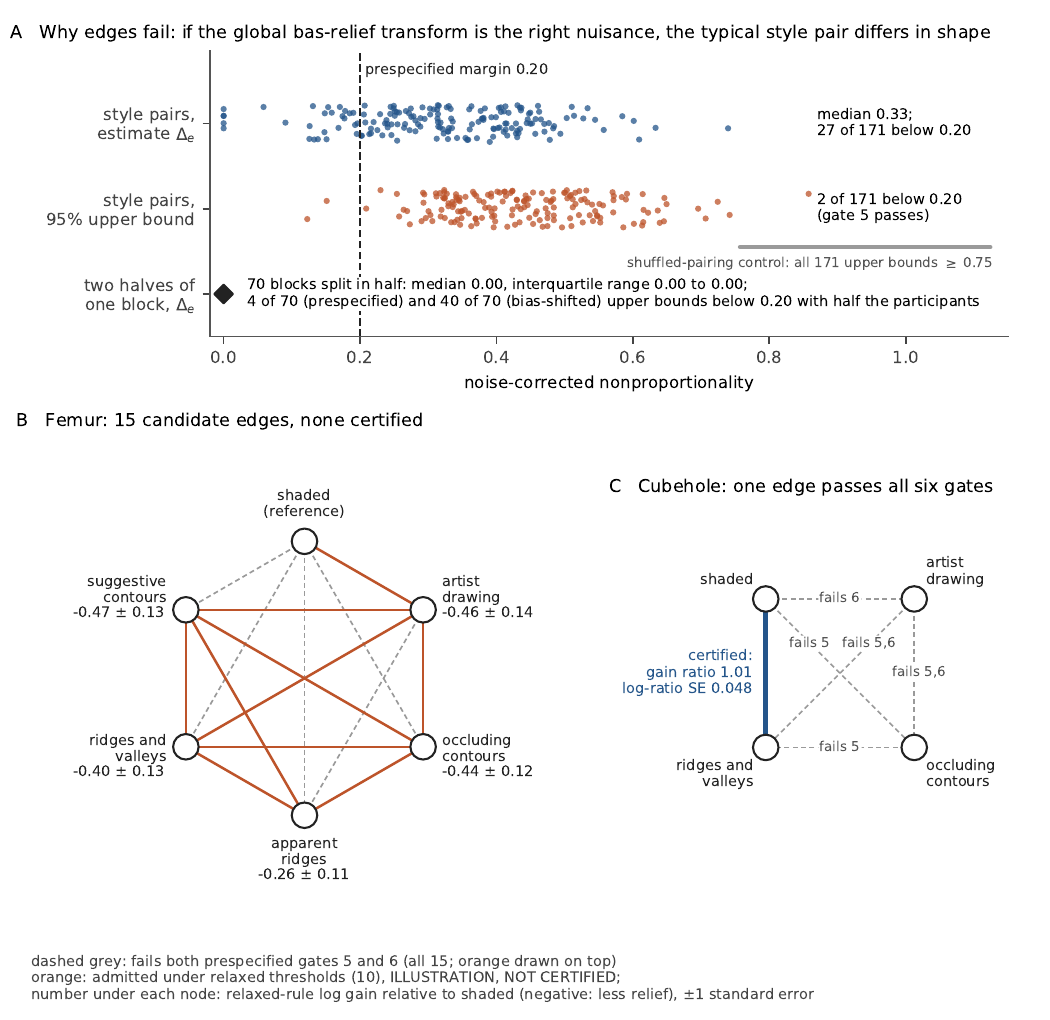}

Figure 3. Certification on the Cole et al.~(2009) depiction-style
blocks. (A) Noise-corrected nonproportionality \(\Delta_e\) (Section
3.4's statistic, noise-corrected as in Section 4) for the 171 candidate
style pairs: point estimates (blue, median 0.33) and 95\% bootstrap
upper bounds (orange), against the prespecified margin 0.20 (dashed
line). Only 2 upper bounds fall below the margin, so gate 5 passes for 2
edges. Two random halves of each of the 70 blocks, proportional by
construction, give a median of 0.00 (diamond; the noise correction
floors at zero); shuffling the probe-point pairing puts every upper
bound at or above 0.75 (grey bar). (B) The femur model's six style
nodes. Each of the 15 candidate edges fails both prespecified gates 5
and 6 (dashed grey; the 10 orange edges are drawn over their dashed
lines). The 10 orange edges are those admitted under thresholds relaxed
after seeing the data (margin 0.30, standard error 0.20); they
illustrate the cycle diagnostic and are not certified. The number under
each node is that style's relaxed-rule log gain relative to the shaded
image, plus or minus one standard error; a negative value means less
relief than the shaded image. (C) The cubehole model, whose shaded-image
to ridges-and-valleys edge is the only one of 171 to pass all six
prespecified gates (gain ratio 1.01, log-ratio standard error 0.048);
every other edge is labelled with the gates it fails. Plotted values:
\texttt{anc/p50\_cole2009\_analysis.py}.

\end{figure}

\textbf{Reasons the edges failed.} For the typical style pair, gate 5
binds for a substantive reason: the noise-corrected nonproportionality
\(\Delta_e\) has median 0.33 across style pairs, against 0.00 for two
random halves of one block, which are proportional by construction
(Figure 3A). The typical pair's participant-averaged residual profiles
are therefore not explained by a single positive gain, provided the
global bas-relief transform is the right nuisance to remove (Section 7).
Reading that as a depiction-style effect on perceived shape further
assumes that the participants who saw each style are comparable. Under
Section 3.1 the styles remain separate nodes. Of the 27 pairs whose
point estimate lies below the margin, only 2 have an upper bound below
it, so for the other 25 gate 5 binds on precision rather than shape. The
prespecified plain percentile bound is wide here: its median upper
bound, 0.43, lies about 0.10 above the median point estimate, 0.33. Gate
6 fails on sample size: the median log-ratio standard error, 0.183,
would need about 3.3 times as many participants per style to reach 0.10,
if the standard error keeps falling as one over the square root of the
sample size.

\hypertarget{post-hoc-analyses}{%
\subsection{4.3 Post hoc analyses}\label{post-hoc-analyses}}

The analyses in this subsection were chosen after seeing the data. Their
counts are descriptive, not prespecified tests, and none of them
certifies an edge.

\textbf{Bias-shifted percentile.} We added a bias-shifted percentile,
the percentile minus the bootstrap estimate of bias, after the gate 4
control of Section 4.4 exposed a defect in the plain percentile. Applied
to every gate, the bias-shifted percentile still passes gates 3 and 4 on
all 171 edges and raises gate 5 passes from 2 to 9. Every plain upper
bound lies at or above its bias-shifted counterpart, so the 2 plain
passes are among the 9. The certified edge is unchanged, because the
other eight fail gate 6. The bias-shifted 95\% lower bound on
\(\Delta_e\) shows the shape difference directly: it lies above the
margin for 120 of 171 style pairs and above zero for 166.

\textbf{Relaxed thresholds: the cycle diagnostic.} Thresholds relaxed
after seeing the data (margin 0.30, standard error 0.20) admit 38 edges
to a descriptive graph, certifying none, and connect the femur model's
six styles with five independent cycles (Figure 3B). There
\(Q_{\rm cycle}=0.81\) on 5 degrees of freedom, but the same data chose
these edges, so by Section 3.3 its chi-square reference does not hold
and the value is descriptive. The prespecification also names a
split-sample version, in which edges are chosen on half the participants
and \(Q_{\rm cycle}\) is computed on the other half. Run under the same
relaxed thresholds, it leaves the femur model no edge, and only the
cubehole model has a cycle (\(Q_{\rm cycle}=0.36\) on 1 degree of
freedom, \(p=0.55\), descriptive because the thresholds were chosen
after seeing the full data).

\textbf{Relaxed thresholds: the transport test.} This example shows how
Section 5.2's transport test reads; it is not a finding about line
drawings, and no relaxed-rule edge is certified. Under the relaxed
rules, the artist's drawing has a lower point-estimate gain than the
shaded image on all four models linking the two. The homogeneity
statistic asks whether that gain ratio is the same across models; it
gives \(p=0.008\), descriptive here because the same data chose the
edges, so the ratio appears to vary across models. The same statistic
for apparent ridges against the shaded image gives \(p=0.61\), and for
ridges and valleys against the shaded image \(p<0.001\).

\hypertarget{controls}{%
\subsection{4.4 Controls}\label{controls}}

\textbf{Controls under the prespecified rules.} Shuffling the
probe-point pairing fails gate 5 on all 171 edges (every upper bound at
or above 0.75). Flipping the sign of one endpoint's profile fails gate 3
on all 171 edges, every one of which passes it unflipped. Split halves
of one block, proportional by construction and each with half the
participants, pass the margin in 4 of 70 blocks. A planted pair that is
exactly proportional at full sample size passes in 19 of 70 blocks. For
each block, two independent sets of simulated participants, as many as
the block has, scatter around the block's mean profile with its
per-point spread, and one set is scaled by a gain of 1.3. A second seed
gives 23 of 70, and scrambling the probe points of one set passes none.
The simulated noise is independent across probe points, unlike real
participants' settings, so real proportional pairs may pass less often
still. Gate 5 therefore fails most truly proportional pairs at these
sample sizes, and neither the 4 of 70 nor the 19 of 70 bounds the number
of truly compatible style pairs. One control exposed a defect: for gate
4, the plain bootstrap percentile excluded zero in 30 of 70 pure-noise
profiles.

\textbf{Controls under the bias-shifted percentile (post hoc).} The
bias-shifted percentile excludes zero in none of the 70 pure-noise
profiles. The split halves pass the margin in 40 of 70 blocks, and their
bias-shifted lower bound on \(\Delta_e\) lies above zero only once,
against 166 of 171 style pairs.

\textbf{Controls under the relaxed thresholds (post hoc).} Multiplying
the shaded block's gradients by 1.5 shifts every other style's log gain
relative to it by a median \(-0.4057\) over 15 contrasts, against the
planted \(-\log 1.5=-0.4055\). A log 1.5 error planted on one cycle edge
pushes \(Q_{\rm cycle}\) past its 95\% critical value in all six models
with cycles, and ratios formed from one set of node gains give
\(Q_{\rm cycle}=0\), as Section 3.3 warns.

\hypertarget{what-the-example-shows}{%
\subsection{4.5 What the example shows}\label{what-the-example-shows}}

On a large published dataset the procedure certified 1 of 171 candidate
links under the prespecified rules. The others failed for two reasons:
too few participants saw each style, and, by the post hoc lower bound
and provided the global bas-relief transform is the right nuisance, most
style pairs differ in shape beyond it.

These results bear on the premise taken from Koenderink et al.~(2001) in
Section 2, that depth measurements agree up to an offset, a depth shear
and a positive gain. Sections 4.2 to 4.4 test the same affine relation,
but across depiction styles within one task. By the post hoc lower
bound, and if the global bas-relief transform is the right nuisance,
most style pairs depart from it by more than the margin (the lower bound
on \(\Delta_e\) exceeds 0.20 for 120 of 171 pairs; only 1 of 171 links
is certified). Styles are a different axis from tasks, so this result
neither confirms nor refutes the cross-task agreement. That agreement
remains untested here, because testing it needs estimates from two
methods on shared stimuli with a joint uncertainty model (Section 1).
The two criteria also differ in stringency. Before noise correction, a
pair of residual profiles passes the 0.20 margin only if the profiles
meet at an angle below about 23 degrees, a squared correlation above
about 0.85 (a necessary condition; Appendix A.8). Koenderink et al.'s
\(R^2\) also credits the image-coordinate terms, so it is at least that
squared profile correlation, and map pairs at the low end of their 0.8
to 1.0 range would fail the margin before noise correction. The
comparison is indicative only, because this section applies the
statistic to noise-corrected depth gradients rather than to depth maps.

\hypertarget{extensions-ordinal-trials-and-a-correlated-error-test}{%
\section{5 Extensions: ordinal trials and a correlated-error
test}\label{extensions-ordinal-trials-and-a-correlated-error-test}}

Two extensions follow. The first (Section 5.1) concerns
probabilistic-comparison data whose discrimination is unknown, and the
second (Section 5.2) gives the test of whether one common slope between
two methods holds across depiction styles.

\hypertarget{ordinal-trials-and-unknown-discrimination}{%
\subsection{5.1 Ordinal trials and unknown
discrimination}\label{ordinal-trials-and-unknown-discrimination}}

For pairwise outcome \(Y_{sijr}\), use

\[
\Pr(Y_{sijr}=1)
=
\frac{\ell_s}{2}
+
(1-\ell_s)
F\!\left[
\gamma_s+d_{sr}(z_i-z_j)
\right].
\]

Here \(s\) indexes the study or session and \(r\) the observer or
replicate within it, \(z_i\) is stimulus \(i\)'s latent score as in
Section 3.6, \(d_{sr}>0\) is discrimination, \(\gamma_s\) is an optional
order/response bias, and \(\ell_s\) is a lapse rate. For any \(q>0\)
(centering \(z\) fixes the origin and is not needed here), the
transformation

\[
z_i'=qz_i,
\qquad
d_{sr}'=\frac{d_{sr}}{q}
\]

leaves every probability unchanged.

A normalization or prior can select a representative but does not make
the scale likelihood-identified. Repeated probabilistic-comparison
studies can sharpen an internal comparison coordinate and its
uncertainty, in the sense of Section 3.6, but with study-specific
unknown discrimination they do not supply an externally transportable
unit. A quantitative edge to another operation requires discrimination
calibrated in compatible units, or a joint quantitative operation that
fixes the unit.

The rank study (van Doorn et al., 2011) illustrates the distinction. Its
stimulus was a close copy, by Anne-Sophie Bonno, of a wash drawing by
Francesco Guardi (1712--1793), an imaginary landscape (p.~728). In the
reported sample, the number of discriminable pictorial-depth layers was
not correlated with total metrical depth range, so the data do not
support a simple positive relation between depth range and depth
resolution. This study-specific observation is separate from the
likelihood invariance proved here. Exact figure-derived values and
session counts are not repeated here; the qualitative statement above is
what the source figure supports.

\hypertarget{correlated-error-test-for-method-transport}{%
\subsection{5.2 Correlated-error test for method
transport}\label{correlated-error-test-for-method-transport}}

For positive gain estimates \(\widehat a_{ph}\) and \(\widehat a_{qh}\)
in style \(h\), the context-invariant gain-ratio null is

\[
H_0:a_{qh}=\kappa a_{ph}
\quad\text{for all }h.
\]

This null is a necessary slope condition for full affine transport, not
a sufficient one. Equal intercept conversions across styles are
additionally necessary (Section 3.7), and this test does not assess
them.

Define

\[
l_h
=
\log\widehat a_{qh}
-
\log\widehat a_{ph}.
\]

If the pair covariance is

\[
S_h
=
\begin{bmatrix}
\sigma_{pp,h} & \sigma_{pq,h}\\
\sigma_{pq,h} & \sigma_{qq,h}
\end{bmatrix},
\]

then

\[
V_h
\approx
\frac{\sigma_{pp,h}}{\widehat a_{ph}^{\,2}}
+
\frac{\sigma_{qq,h}}{\widehat a_{qh}^{\,2}}
-
\frac{2\sigma_{pq,h}}
{\widehat a_{ph}\widehat a_{qh}}.
\]

This is the delta-method variance of the estimated log ratio. Its
covariance term can either raise or lower the uncertainty, so \(V_h\)
cannot be recovered from marginal standard errors alone.

If styles are independent, the diagonal-covariance special case gives

\[
\widehat\theta
=
\frac{\sum_h l_h/V_h}
{\sum_h1/V_h},
\qquad
Q_{\rm diag}
=
\sum_h
\frac{(l_h-\widehat\theta)^2}{V_h}
\overset{H_0}{\approx}
\chi^2_{n_H-1}.
\]

Here \(n_H\) is the number of styles, \(\widehat\theta\) is the
inverse-variance weighted common log ratio and \(Q_{\rm diag}\) measures
residual between-style heterogeneity around that common value.

If styles share participants, stimuli, masks, or preprocessing, let

\[
l=(l_1,\ldots,l_{n_H})^\top
\]

and use the full covariance

\[
\Sigma_l=\operatorname{Cov}(l).
\]

Under

\[
l=\theta\mathbf1+\varepsilon,
\]

\[
\widehat\theta_{\rm GLS}
=
\frac{\mathbf1^\top\Sigma_l^{-1}l}
{\mathbf1^\top\Sigma_l^{-1}\mathbf1},
\qquad
Q_{\rm GLS}
=
(l-\widehat\theta_{\rm GLS}\mathbf1)^\top
\Sigma_l^{-1}
(l-\widehat\theta_{\rm GLS}\mathbf1).
\]

These are the same pooled-ratio and heterogeneity calculations with the
independent style weights replaced by the full cross-style covariance.

Under standard regularity and known-covariance conditions,
\(Q_{\rm GLS}\) is approximately

\[
\chi^2_{n_H-1}.
\]

Replacing \(\Sigma_l\) by a covariance estimate from the same finite
sample does not preserve that reference distribution in general. An
exact Hotelling-type \(F\) calibration requires multivariate normality
and the corresponding independent Wishart covariance structure;
arbitrary shared-participant, shared-stimulus, or preprocessing
dependence does not supply it. Otherwise use a justified finite-sample
correction or a parametric or bootstrap calibration that reproduces
covariance estimation.

Near zero, use direct constrained generalized least squares, profile
likelihood, or a calibrated resampling procedure instead of log-Wald
inference. The heterogeneity statistic tests a common ratio; a large
\(p\)-value does not establish equivalence. The empirical active-edge
decision additionally requires the proportionality and covariance gates
(gates 5 and 6), made operational in Section 3.4. None of the published
aggregate results examined here provides all the paired covariance and
bridge-specific information that the certification rule in Section 3.4
requires to certify a new cross-method active edge.

\hypertarget{discussion}{%
\section{6 Discussion}\label{discussion}}

\hypertarget{implications-for-measurement-and-design}{%
\subsection{6.1 Implications for measurement and
design}\label{implications-for-measurement-and-design}}

The opening case can now be read precisely. Observers whose depth maps
correlate highly but carry different scale factors, as in the
relative-size study, agree closely in shape and differ in gain. Theorem
1 says, under its assumptions, when the active overlaps alone place
those gains in one relative unit: exactly when the observers' blocks are
joined by a connected graph of active overlaps; other evidence, such as
the anchors of Lemma 1, is a separate route. Lemma 1 adds the stronger
condition under which they also share an origin. Neither result says
that the observers see the same depth. The account can also fail on
data. In a study that records two methods for the same participants and
pictures, a proposed cross-method link fails if the prespecified upper
bound on \(\Delta_e\) for its shared contrasts exceeds the equivalence
margin. The public Cole et al.~(2009) settings allow only the
within-method version of that test, across participants and depiction
styles, and there only 1 of 171 candidate links passed (Section 4). No
dataset examined here supplies the two-method estimates on shared
stimuli, with joint uncertainty information, that the cross-method test
needs.

Transport claims require more than generic agreement. Ordinal order,
quantitative configuration, rating response, and viewing layout remain
distinct. A common relative unit arises from connected active overlaps
after nuisance quotienting; sharing the label ``depth'' does not create
one.

Landy et al.~(1995, pp.~392, 394) show that cue outputs may require
additional parameters before they are commensurate. Their promotion
operation differs from calibration and does not establish cross-task
measurement transport.

The graph also offers an experimental-design principle. If two
scientifically compatible measurement blocks lie in different
components, a validated nonzero bridge can reduce the number of
independent units. A more precise measurement within one existing
component improves precision but cannot identify the relative unit
between components. After operation type, nuisance equivalence, viewing
condition, and construct compatibility have been checked, candidate
bridges can be prioritized by component reduction, expected variance
reduction for prespecified contrasts, overidentification contribution,
and deletion robustness. Given the graph, its weights and a prespecified
contrast, three of those four, component reduction, overidentification
contribution and deletion robustness, are exact quantities rather than
directions of preference; the fourth, expected variance reduction, is
exact once the new bridge's information weight is assumed. Once the
graph is connected and edge errors are independent, the precision of a
named contrast is its effective resistance. The one-edge reduction
formula of Section 3.3 can then rank proposed edges within a connected
part of the graph by their anticipated precision before any of them is
run. The ranking is conditional: such an edge adds that precision only
if its data later pass gates 3 to 6, which the formula cannot guarantee.
For the stronger common-coordinate target and algebraically generic
anchor values, the partition-deficit formula of Section 3.6 says exactly
how many affine directions a deficient design is missing. It also says
which splits of the charts (measurement blocks, not tasks) are
responsible; a maximizing partition need not be unique. The
leave-one-anchor-out criterion says what margin of crossing anchors lets
the design survive losing one. A realized anchor assignment must still
pass the matrix-rank check, as the \(K_4\) example of Section 3.6 shows.
A missing unit therefore names the next experiment.

The empirical literature motivates the restrictions. Relative-size depth
has no natural origin or unit in the studied operation. Rank resolution
and pointing-derived range fail to support a simple positive
range-resolution relation in the reported sample. Koenderink and van
Doorn's gauge-field analysis includes a global transformation with
additive image-position terms and multiplicative depth scaling, as well
as more general space-varying gauge fields. The nuisance basis

\[
Q=[\mathbf1,x,y]
\]

used here captures only the additive image-position component and is a
declared approximation, not a complete account of that framework. These
source-specific empirical findings and the gauge analysis support typed,
condition-specific measurement blocks instead of an assumed universal
depth variable.

The gauge and graph results concern measurement identification, not a
complete generative account of pictorial perception. The Gestalt review
(Wagemans et al., 2012, ``A century of Gestalt psychology in visual
perception'') links grouping and figure-ground assignment to perceived
surface layout, while Bayesian object-perception work (Kersten,
Mamassian, and Yuille, 2004, p.~271) analyzes ambiguity through image
evidence and prior knowledge. These accounts may motivate future
experimental contrasts, but they do not supply a calibrated bridge
between the operations studied here.

The published examples cover a narrow region of depiction and
presentation conditions. The title and claims therefore concern methods
and conditions, not artworks themselves; no claim is made about any
particular artwork or artistic style, and the depiction styles of
Section 4 are computer renderings of 3-D models. Testing broader
art-historical scope requires a separate empirical program.

\hypertarget{relation-to-prior-work}{%
\subsection{6.2 Relation to prior work}\label{relation-to-prior-work}}

The search for antecedents, run in September and October 2026 with web
and literature-database queries, covered psychometric linking and
equating, sensor-network and image calibration, reviewer and rater
calibration, synchronization on graphs, generic matrix completion and
rigidity, spanning-tree packing, econometric identification, and
comparisons of pictorial relief across tasks. The paper's contribution
is a typed framework. It combines heterogeneous psychological methods
with incomplete overlap, specified nuisance equivalence, exceptional
anchor geometry, weak conditioning, falsifiability, and distinct levels
of context transport. We do not claim novelty for affine linking,
calibration, graph synchronization, two-tree combinatorics, network
least-squares linking of many scales, or the finding that depth tasks
agree once an affine ambiguity is removed. Theorem 1 in particular is
the standard identifiability statement for synchronization on a graph
over the real numbers under addition (Singer, 2011); Appendix A.1 says
so where it proves the theorem.

Three earlier lines of work already did parts of this, and the paper's
claims are narrowed to what they leave open. The pictorial-relief line,
Koenderink, van Doorn, Kappers, and Todd (2001), is described in Section
2, where this paper takes its nuisance basis from their affine
regression.

In psychometrics, Haberman (2009) links many separately calibrated test
forms simultaneously by least squares on log scale factors over pairwise
common items, generalizing log-mean-mean linking. His estimates exist
exactly when the forms are connected through shared items (pp.~3-5), the
counterpart of Theorem 1, and the estimator of Section 3.3 is a
covariance-weighted relative of his. Battauz (2013) averages the
equating coefficients that link the same two forms along different
paths. The weights minimize the variance of the average, the standard
errors allow for paths sharing links (pp.~466-468), and Battauz proposes
testing whether different paths give different coefficients (p.~477).
Battauz (2017) estimates the coefficients of all forms at once, with
asymptotic standard errors (pp.~610-613). Network estimation of gains,
efficient path weighting and path-disagreement checks are therefore
borrowed, not new. None of these authors states a condition for fixing
unit and origin from the same equations. Outside psychometrics, Tan, Wu,
Bei, and Xu (2021) do state one, for noiseless peer-review scores in
which each reviewer applies an increasing linear map with its own scale
and offset (p.~27071). They prove that every assignment and every
admissible set of scores is recovered up to one global positive affine
map exactly when reviewer groups that share at least two items can be
merged, repeatedly, into one group covering every item (Theorem 3.4,
p.~27073), because two shared items fix unit and origin together. Their
condition guarantees recovery for all distinct anchor values and is
strictly stronger than two edge-disjoint spanning trees: the well-spread
\(K_4\) design of Section 3.6 has two trees but no pair of charts
sharing two anchors. This paper's additions are narrower: typed gates
that decide whether an overlap may enter the network at all; a
Sherman--Morrison score of the variance that an edge the network does
not yet have would remove from one prespecified contrast; and the
generic common-coordinate analysis of Section 3.6, which also covers
designs such as \(K_4\) that their condition does not. Haberman's
procedure fixes each form's unit first, from item discrimination
parameters, and then needs only connectivity to fix the origins.
Pictorial-depth methods have no such parameter, so unit and origin must
come from the same equations. That double demand raises the requirement
from one spanning tree to two, the generic counterpart of Tan et al.'s
rule that merged groups share two items.

Van der Linden and Barrett (2016) derive the form of item response
theory linking functions for monotone, continuous response models from
the models' lack of identifiability (pp.~650, 661). They then show that
the linking parameters are identified from the parameter values of items
or persons common to the two calibrations; their 2017 erratum supplies
the missing separability step in the proof of their Theorem 3. Bach
(2025, p.~1) formalizes calibration against experimentally predicted
standard scores, and Meredith (1993, p.~528) treats measurement
invariance through conditional response distributions. Research on
direction and rigidity constraints gives combinatorial background,
through two sources with different roles. Whiteley (1988) proves the
matroid-union result that Section 3.6's Proposition 2 specializes, and
is cited there for it. Servatius and Whiteley (1999, p.~136) belong to
the computer-aided-design literature on constraining plane
configurations by directions and lengths; that literature is adjacent
background, not the source of the matroid-union statement used here. A
self-contained determinant argument below proves the generic two-tree
result for the pairwise affine-calibration design, where generic means
true for all anchor values outside a set of Lebesgue measure zero.

Mancinelli and Bach (2026, p.~1) address inference and decision-making
when several measurement methods compete by retrodictive validity
against experimentally predicted standards. That framework compares
methods intended to measure the same experimentally manipulated latent
variable. The present problem is complementary: it asks whether typed
methods with incomplete shared observations admit a common unit or
coordinate at all, before their accuracy can be ranked on one scale.

\hypertarget{numerical-source-checks}{%
\subsection{6.3 Numerical source checks}\label{numerical-source-checks}}

We attempted three numerical source checks against the primary articles
and completed two. The rank-study figure values (van Doorn et al., 2011)
support the qualitative statement of Section 5.1. Zhao et al.~(2024)
report in prose a mean engraving-to-painting ratio of 0.71 (p.~10) for
the standard deviation of z-scored ratings, their measure of perceived
range. The ratio is taken over twenty sessions covering five perceptual
attributes (three-dimensionality, glossiness, convincingness, smoothness
and softness). Depth-like content is therefore one attribute of five,
and the ratio is not a measure of how deep either medium looks. That
prose value does not match their own Table 3 (p.~11): the twenty ratios
formed from its printed standard deviations have mean 0.88, and the
ratio of the mean standard deviations is also 0.88. No average of the
ratios can be 0.71, because two consecutive sentences give their range
as 0.69 to 1.14 and place 3 of the 20 at or above 1, which forces a mean
of at least 0.73; 0.71 is the mean correlation reported in the preceding
paragraph. The recalculated 0.88 documents the source inconsistency
only; no inferential or calibration claim in this paper depends on it. A
secondary reciprocal-regression derivation from the point-configuration
tables (Wagemans, van Doorn, and Koenderink, 2011, ``Measuring 3D point
configurations in pictorial space'', Tables 1-2, pp.~90-91) remains
excluded because the primary text does not fully specify the
directional-regression convention and observation pairing required for
that calculation.

\hypertarget{limitations}{%
\section{7 Limitations}\label{limitations}}

Section 4 is this paper's only analysis of participant-level records,
and it covers depiction styles within one method; method-by-style tests
still await raw records.

Priors can regularize the flat direction of the ordinal likelihood,
along which scale and discrimination trade off (Section 5.1), but cannot
remove its likelihood invariance.

The affine nuisance basis is a declared approximation. A poor basis can
erase signal or leave structured residual differences; an acceptable
gauge must be justified for the task and checked on held-out contrasts.
Section 4's conclusion that most style pairs differ in shape assumes
that the global bas-relief transform is the right one to remove. Cole et
al.~(2009, p.~8) note, following earlier work, that local bas-relief
fitting may model perceived shape more accurately. The split-half
control cannot tell a shape difference from a local gauge that the
global basis omits, because both halves share style and basis.

The source-grounded examples use few artworks and observers. Their
observed patterns do not establish population-wide stability or
stability across artistic traditions.

The rank checks of Appendix B.4 (S1) are algebraic. Finite-sample
coverage near singular or positivity boundaries remains outside this
paper's simulation claim.

The correlated-error simulation uses known equicorrelated Gaussian
covariance. It does not validate covariance estimation under unmodeled
heterogeneity or dependence.

The secondary reciprocal-regression derivation remains excluded because
the primary source does not fully specify the regression convention and
observation pairing needed for that calculation. The other two source
checks do not alter the paper's substantive claims.

\hypertarget{conclusion}{%
\section{8 Conclusion}\label{conclusion}}

The paper opened with observers who agree about what lies in front in a
picture yet report different amounts of depth. Their depth amounts can
be compared only when the blocks that produced them share a unit.

Treat each measurement block as a node and each overlap that passes the
six gates as a link. If the graph is connected and the declared model
holds, every block's gain is fixed in one common relative unit (Theorem
1), as linked test forms are in psychometrics. A shared unit is not a
shared depth coordinate: fixing origins too requires the rank criterion
of Lemma 1, under an assumed compatibility condition and its nullity
condition. A design with only enough links to fix the units leaves
nothing to test fit, and ordinal judgments with unknown discrimination
fix no further scale freedom.

The evidence is narrow. Section 4 certifies one link, between two
depiction styles measured with one probe method; no link between methods
is yet certified. A future paired-record analysis can test such links
with declared nuisance effects, full covariance, an equivalence margin
and a gate 6 information minimum fixed in advance, in a design that
stays identified if any one anchor is removed.

The practical message is that a common scale can be designed rather than
hoped for, because each kind of missing unit leaves a visible trace and
calls for its own next measurement. A graph in separate pieces, or
anchors that leave some blocks' origin or unit free to move relative to
the rest, means a unit is missing, and the researcher restores it by
collecting overlaps that join the pieces or cross the deficient split.
For generic anchor values a poor split of the blocks explains the gap,
but particular anchor values can produce one even when every split meets
the condition (the four-block \(K_4\) example, Section 3.6). Blocks
joined only through weak, noisy overlaps share a unit that is identified
but imprecise; more overlaps along that path sharpen it. An untested
viewing context leaves transport open until someone measures in it.
Researchers comparing observers, tasks or styles can thus read off which
new data would put their numbers on one scale, provided those data pass
the gates and fit a compatible, nondegenerate model.

\hypertarget{data-and-code-availability}{%
\section{Data and code availability}\label{data-and-code-availability}}

A reproduction package accompanies this paper as arXiv ancillary files
(folder \texttt{anc/}, run with \texttt{anc/run\_all.sh}), containing
the graph enumeration, exact S1 rank checks, ordinal ridge check,
covariance simulation, affine-gauge simulation, the S2 generator and
fitter, the fixed random seeds written into those scripts, and a
manifest writer (\texttt{anc/p50\_write\_manifest.py}) that records
seeds, environment versions and output hashes when the package is run.
The printed numbers were produced under Python 3.8.8, NumPy 1.20.1,
SciPy 1.6.2 and matplotlib 3.3.4. The code behind Appendix B.3's
original run was not preserved. A replacement script in the package
(\texttt{anc/p50\_covariance\_sim.py}) reproduces all six of its
rejection rates. Each reproduced rate lies within 2.2 standard errors of
the printed rate and of any value that rounds to it at four decimals.
The standard error here is that of the difference between two
independent runs of 200,000 replicates each. Appendix B.4 reports the
numbers that the shipped S2 generator and fitter produce. No
participant-level records are included in the package. Section 4
reanalyses the public attitude-probe (gauge-figure) dataset of Cole et
al.~(2009), version 1.0 (\url{https://lineshape.cs.princeton.edu/}, GNU
General Public License v3), which is not redistributed here. The script
\texttt{anc/p50\_cole2009\_fetch.sh} downloads it and checks its SHA-256
checksum
(e1449f85e17a238f67a1a30058b1d3844b57690edf2e663eb20e9196373c2224), and
\texttt{anc/p50\_cole2009\_analysis.py} reproduces every number in that
section, with its fixed seed, in under three minutes
(\texttt{bash\ anc/p50\_cole2009\_fetch.sh}, then
\texttt{python3\ anc/p50\_cole2009\_analysis.py}), apart from the second
implementation and the planted-pair control. Those come from
\texttt{anc/p50\_cole2009\_reimpl.py} (run with \texttt{decide}, then
\texttt{compare}) and \texttt{anc/p50\_cole2009\_planted.py}. The
prespecified thresholds were fixed in advance in
\texttt{anc/p50\_cole2009\_prespecification.md}, committed to the
project repository on 2026-10-15, before the first commit of the
analysis script; no external registry timestamp exists. The relaxed
thresholds of Section 4 were not fixed in advance. A stable external
archive identifier has not yet been assigned.

\hypertarget{use-of-ai-tools}{%
\section{Use of AI tools}\label{use-of-ai-tools}}

AI tools assisted under the author's direction. ChatGPT (OpenAI,
including its Codex agent) and Claude (Anthropic) wrote analysis,
simulation and figure code to the author's specification and assisted in
editing the final paper. Perplexity was used for literature search and
citation verification. ChatGPT (OpenAI), Claude (Anthropic) and Gemini
(Google) were used for adversarial review and independent mathematical
checking, and Wolfram Mathematica for checking derivations.

\hypertarget{references}{%
\section{References}\label{references}}

Wagemans, J., van Doorn, A. J., and Koenderink, J. J. (2011). Measuring
3D point configurations in pictorial space. \emph{i-Perception}, 2,
77-111. \url{https://doi.org/10.1068/i0420}

van Doorn, A. J., Koenderink, J. J., and Wagemans, J. (2011). Rank order
scaling of pictorial depth. \emph{i-Perception}, 2, 724-744.
\url{https://doi.org/10.1068/i0432aap}

Wagemans, J., van Doorn, A. J., and Koenderink, J. J. (2011). Pictorial
depth probed through relative sizes. \emph{i-Perception}, 2, 992-1013.
\url{https://doi.org/10.1068/i0474}

van Doorn, A. J., Koenderink, J. J., Leyssen, M. H. R., and Wagemans, J.
(2012). Interaction of depth probes and style of depiction.
\emph{i-Perception}, 3, 528-540. \url{https://doi.org/10.1068/i0500}

Zhao, Y., Stumpel, J., de Ridder, H., and Wijntjes, M. W. A. (2024).
Material perception across different media - comparing perceived
attributes in oil paintings and engravings. \emph{i-Perception}, 15(4),
1-21. \url{https://doi.org/10.1177/20416695241261140}

Koenderink, J. J., and van Doorn, A. J. (2012). Gauge fields in
pictorial space. \emph{SIAM Journal on Imaging Sciences}, 5, 1213-1233.
\url{https://doi.org/10.1137/120861151}

Landy, M. S., Maloney, L. T., Johnston, E. B., and Young, M. (1995).
Measurement and modeling of depth cue combination: In defense of weak
fusion. \emph{Vision Research}, 35, 389-412.
\url{https://doi.org/10.1016/0042-6989(94)00176-M}

Stevens, S. S. (1946). On the theory of scales of measurement.
\emph{Science}, 103(2684), 677-680.
\url{https://doi.org/10.1126/science.103.2684.677}

Wagemans, J., Elder, J. H., Kubovy, M., Palmer, S. E., Peterson, M. A.,
Singh, M., and von der Heydt, R. (2012). A century of Gestalt psychology
in Visual Perception: I. Perceptual Grouping and Figure-Ground
Organization. \emph{Psychological Bulletin}, 138, 1172-1217.
\url{https://doi.org/10.1037/a0029333}

Kersten, D., Mamassian, P., and Yuille, A. (2004). Object perception as
Bayesian inference. \emph{Annual Review of Psychology}, 55, 271-304.
\url{https://doi.org/10.1146/annurev.psych.55.090902.142005}

van der Linden, W. J., and Barrett, M. D. (2016). Linking item response
model parameters. \emph{Psychometrika}, 81, 650-673.
\url{https://doi.org/10.1007/s11336-015-9469-6}

van der Linden, W. J., and Barrett, M. D. (2017). Erratum to: Linking
item response model parameters. \emph{Psychometrika}, 82, 273.
\url{https://doi.org/10.1007/s11336-016-9551-8}

Bach, D. R. (2025). Experiment-based calibration in psychology:
Foundational and data-generating model. \emph{Journal of Mathematical
Psychology}, 127, 102950.
\url{https://doi.org/10.1016/j.jmp.2025.102950}

Mancinelli, F., and Bach, D. R. (2026). Experiment-based calibration:
Inference and decision-making. \emph{Behavior Research Methods}, 58,
article 186. \url{https://doi.org/10.3758/s13428-026-03042-9}

Meredith, W. (1993). Measurement invariance, factor analysis and
factorial invariance. \emph{Psychometrika}, 58, 525-543.
\url{https://doi.org/10.1007/BF02294825}

Koenderink, J. J., van Doorn, A. J., Kappers, A. M. L., and Todd, J. T.
(2001). Ambiguity and the `mental eye' in pictorial relief.
\emph{Perception}, 30, 431-448. \url{https://doi.org/10.1068/p3030}

Haberman, S. J. (2009). Linking parameter estimates derived from an item
response model through separate calibrations. \emph{ETS Research Report
Series}, 2009(2), ETS RR-09-40.
\url{https://doi.org/10.1002/j.2333-8504.2009.tb02197.x}

Battauz, M. (2013). IRT test equating in complex linkage plans.
\emph{Psychometrika}, 78, 464-480.
\url{https://doi.org/10.1007/s11336-012-9316-y}

Battauz, M. (2017). Multiple equating of separate IRT calibrations.
\emph{Psychometrika}, 82, 610-636.
\url{https://doi.org/10.1007/s11336-016-9517-x}

Bradley, R. A., and Terry, M. E. (1952). Rank analysis of incomplete
block designs: I. The method of paired comparisons. \emph{Biometrika},
39, 324-345. \url{https://doi.org/10.2307/2334029}

Ghosh, A., Boyd, S., and Saberi, A. (2008). Minimizing effective
resistance of a graph. \emph{SIAM Review}, 50, 37-66.
\url{https://doi.org/10.1137/050645452}

Jiang, X., Lim, L.-H., Yao, Y., and Ye, Y. (2011). Statistical ranking
and combinatorial Hodge theory. \emph{Mathematical Programming}, 127,
203-244. \url{https://doi.org/10.1007/s10107-010-0419-x}

Singer, A. (2011). Angular synchronization by eigenvectors and
semidefinite programming. \emph{Applied and Computational Harmonic
Analysis}, 30, 20-36. \url{https://doi.org/10.1016/j.acha.2010.02.001}

Servatius, B., and Whiteley, W. (1999). Constraining plane
configurations in computer-aided design: Combinatorics of directions and
lengths. \emph{SIAM Journal on Discrete Mathematics}, 12, 136-153.
\url{https://doi.org/10.1137/S0895480196307342}

Cole, F., Sanik, K., DeCarlo, D., Finkelstein, A., Funkhouser, T.,
Rusinkiewicz, S., and Singh, M. (2009). How well do line drawings depict
shape? \emph{ACM Transactions on Graphics}, 28(3), Article 28, 1-9.
\url{https://doi.org/10.1145/1531326.1531334}

Belhumeur, P. N., Kriegman, D. J., and Yuille, A. L. (1999). The
bas-relief ambiguity. \emph{International Journal of Computer Vision},
35(1), 33-44. \url{https://doi.org/10.1023/A:1008154927611}

Nash-Williams, C. St.~J. A. (1961). Edge-disjoint spanning trees of
finite graphs. \emph{Journal of the London Mathematical Society}, s1-36,
445-450. \url{https://doi.org/10.1112/jlms/s1-36.1.445}

Tutte, W. T. (1961). On the problem of decomposing a graph into \(n\)
connected factors. \emph{Journal of the London Mathematical Society},
s1-36, 221-230. \url{https://doi.org/10.1112/jlms/s1-36.1.221}

Whiteley, W. (1988). The union of matroids and the rigidity of
frameworks. \emph{SIAM Journal on Discrete Mathematics}, 1, 237-255.
\url{https://doi.org/10.1137/0401025}

Tan, S., Wu, J., Bei, X., and Xu, H. (2021). Least square calibration
for peer reviews. \emph{Advances in Neural Information Processing
Systems}, 34, 27069-27080.
\url{https://proceedings.neurips.cc/paper_files/paper/2021/file/e354fd90b2d5c777bfec87a352a18976-Paper.pdf}.
Extended version with proofs: arXiv:2110.12607.

\hypertarget{appendix-a-proofs-and-the-sources-of-borrowed-steps}{%
\section{Appendix A: Proofs and the sources of borrowed
steps}\label{appendix-a-proofs-and-the-sources-of-borrowed-steps}}

\hypertarget{a.1-theorem-1}{%
\subsection{A.1 Theorem 1}\label{a.1-theorem-1}}

\textbf{Proof.} Orient the active edges and write

\[
\alpha_v=\log a_v.
\]

Since both endpoints observe the same nonzero residual configuration,
each population edge ratio \(\rho_e>0\) supplies

\[
\alpha_v-\alpha_u=r_e,
\qquad
r_e=\log\rho_e.
\]

Taking logarithms turns a multiplicative gain ratio into an additive
difference of node quantities, so every active edge contributes exactly
one difference equation. Throughout, \(\rho_e\) denotes the positive
proportionality ratio on edge \(e\) and \(r_e\) its logarithm; the
network estimator of Section 3.3 works with \(r_e\), and the
proportionality residual of Section 3.4 works with \(\rho_e\).

The Jacobian of these equations is the transpose of the oriented
incidence matrix, which has rank \(n-c(G^*)\). Its null space consists
of vectors constant within components, corresponding to one
multiplicative scale action per component. No observation links the
scales of different components. The word ``dimension'' refers to the
residual scale action after quotienting the declared nuisance gauge;
that quotient removes the additive nuisance directions of the raw
parameters. \(\square\)

Theorem 1 contains no new mathematics. Once logarithms turn the gain
ratios into differences, the problem is synchronization on a graph over
the real numbers under addition, and Theorem 1 is that problem's
identifiability statement. Singer (2011) states the criterion for the
circle group. With exact measurements, the synchronization problem is
solvable exactly when the measurement graph is connected, the solution
is unique up to one additive constant, and offsets between different
components cannot be determined at all (p.~20). The same paper names the
additive real group as a case of its framework, in the form
\(t_i-t_j=t_{ij}\) with \(t_i\in\mathbb R\), and notes that the
Gaussian-error version is solved by least squares (p.~35). Our rank
statement \(n-c(G^*)\) is that criterion counted rather than described,
since it is the rank of a graph incidence matrix.

\hypertarget{a.2-proposition-1}{%
\subsection{A.2 Proposition 1}\label{a.2-proposition-1}}

\textbf{Proof.} The criterion is minimized by
\(\widehat\alpha_0=J^{+}B\Sigma_r^{-1}r\), the solution orthogonal to
the null space of \(B^\top\), and
\(\operatorname{Cov}(\widehat\alpha_0)=J^{+}B\Sigma_r^{-1}\Sigma_r\Sigma_r^{-1}B^\top J^{+}=J^{+}JJ^{+}=J^{+}\).
Any other admissible gauge changes \(\widehat\alpha\) only by a vector
in that null space, the vectors constant within each component, so its
full node covariance can differ from \(J^{+}\); an estimable \(c\) is
orthogonal to every such vector, so \(c^\top\widehat\alpha\) and its
variance \(c^\top J^{+}c\) do not depend on the gauge. This gives the
first display. Under independent edge errors \(J=BWB^\top=L_W\) is the
weighted graph Laplacian, and the pair contrast \(c=\iota_v-\iota_u\)
gives the standard Laplacian-pseudoinverse expression for weighted
effective resistance. \(\square\)

The identity is the standard Laplacian and effective-resistance geometry
of weighted networks. Ghosh, Boyd, and Saberi (2008) give the pair
formula in exactly the form used here, writing the effective resistance
between two nodes as \((e_i-e_j)^\top G^{\dagger}(e_i-e_j)\) for the
weighted Laplacian \(G\) and its pseudoinverse \(G^{\dagger}\) (their
Eq. (8), p.~42). Ghosh, Boyd, and Saberi also give the estimation
reading of the first display. They interpret total effective resistance
as an A-optimal experiment-design objective, in which a parameter vector
estimated from noisy linear measurements along edge vectors has error
covariance equal to the pseudoinverse of the weighted sum of those
edges' outer products (pp.~49-50). The application here differs in one
respect. It comes only after typed edge certification and nuisance
reduction, so every edge of the measured network has passed the gates of
Figure 1 and Section 3.1, made operational in Section 3.4.

\hypertarget{a.3-the-one-edge-score-and-earlier-edge-derivatives}{%
\subsection{A.3 The one-edge score and earlier edge
derivatives}\label{a.3-the-one-edge-score-and-earlier-edge-derivatives}}

This appendix compares the one-edge variance reduction \(\Delta V(c)\)
of Section 3.3 with the edge derivatives of Ghosh, Boyd, and Saberi
(2008). Ghosh, Boyd, and Saberi (2008) differentiate the \emph{total}
effective resistance, summed over every node pair, with respect to the
conductances of edges the graph \emph{already has} (their Eq. (20),
p.~46), and they state that they do not need the corresponding per-pair
derivative. The quantity needed here is different in all three respects:
it is exact rather than infinitesimal, it concerns one prespecified
contrast rather than the sum over all pairs, and it scores an edge the
graph does not yet have.

\hypertarget{a.4-the-cycle-statistic-and-the-graph-hodge-decomposition}{%
\subsection{A.4 The cycle statistic and the graph-Hodge
decomposition}\label{a.4-the-cycle-statistic-and-the-graph-hodge-decomposition}}

Section 3.3 fits the node log gains by a covariance-weighted projection
of the observed edge log-ratios, and \(Q_{\rm cycle}\) measures the part
of the edge field that the projection leaves over. For diagonal
covariance this construction is the gradient-versus-remainder half of
the weighted graph-Hodge decomposition, which splits an edge flow into a
gradient part and the part orthogonal to it. Jiang, Lim, Yao, and Ye
(2011) establish that split as a weighted least-squares projection of an
edge flow onto the gradient subspace. They show that the fitted score
solves a graph-Laplacian normal equation whose minimum-norm solution
uses the Laplacian's Moore-Penrose pseudoinverse, and that the residual
carries the inconsistency (their Theorem 3, pp.~227-228). Their
decomposition has three parts rather than two, because it further
separates that residual into a locally cyclic curl component and a
globally cyclic harmonic component, and this finer split requires the
triangles of the comparison graph's clique complex (their Theorem 2,
p.~224). This paper uses only the coarse two-part split, so it neither
needs nor claims the triangle structure, and it does not inherit the
local-versus-global reading of inconsistency that the refinement
supplies. Diagonal covariance guarantees the graph reading.

If edge errors correlate, the fit stays orthogonal to the residual and
the projection reading survives. The normal-equation matrix
\(J=B\Sigma_r^{-1}B^\top\) keeps zero row sums, but it is the Laplacian
of the measured graph with nonnegative weights only when every
off-diagonal entry is nonpositive and every pair of blocks that no edge
joins has a zero entry. Correlation can violate either condition; it
does not have to, since a change in edge precision confined to the cycle
space, the edge flows that \(B\) sends to zero, leaves \(J\) unchanged.
Independent errors are therefore sufficient for the graph reading, not
necessary, and outside them the graph reading has to be checked entry by
entry. The gain model itself is an additive synchronization problem on a
graph once logarithms are taken, as Appendix A.1 sets out against Singer
(2011). Naming that structure does not weaken the diagnostic's limits.

\hypertarget{a.5-lemma-1-and-the-two-tree-condition}{%
\subsection{A.5 Lemma 1 and the two-tree
condition}\label{a.5-lemma-1-and-the-two-tree-condition}}

\textbf{Proof.} The positive affine action is

\[
z_i'=kz_i+t,
\qquad
\lambda_m'=k\lambda_m,
\qquad
\nu_m'=k\nu_m+t,
\qquad
k>0.
\]

At any solution \(\eta_0\), its scale direction \(\eta_0\) and common
translation direction

\[
\tau=(\mathbf 1_N,\,0,1,\,0,1,\ldots,0,1),
\]

that is, 1 in every \(z_i\) and \(\nu_m\) position and 0 in every
\(\lambda_m\) position, are independent elements of \(\ker K\). If the
kernel is exactly their span, every solution with positive slopes has
this form, and positivity forces \(k>0\). If the kernel is larger, a
sufficiently small step along any third null direction preserves
positive slopes and gives a solution outside the global affine orbit.
This establishes global identification for compatible population charts,
not a claim about finite-sample estimator stability.

The two-tree statement of Section 3.6 can be checked directly from
\(H(z)=[D(z)A\mid A]\). To see the calibration determinant directly,
delete one column from each copy of \(A\), select \(2M-2\) edge rows,
and expand the resulting determinant across its first \(M-1\) columns.
Each term chooses \(M-1\) edges for the first block and the
complementary edges for the second; its coefficient is the product of
the two incidence minors, nonzero exactly when both edge sets are
spanning trees. Different allocations of the selected edge rows to the
two incidence blocks carry different monomials

\[
\prod_{e\in T}z_e,
\]

so nonzero terms cannot cancel identically. For generic anchors, a
two-tree subgraph is therefore equivalent to

\[
\operatorname{rank}H=2M-2.
\]

This determinant argument establishes the calibration rank result
directly, without routing it through a rigidity theorem. The matrix
pattern it works on is nonetheless a known one. Whiteley (1988) studies
the general \(k\)-frame on a multigraph, an \(|E|\) by \(k|V|\) matrix
whose row for an edge carries \(k\) distinct indeterminates under one
endpoint and their negatives under the other (his Definition 2.2,
p.~238). Our \(H(z)=[D(z)A\mid A]\) is that matrix at \(k=2\) after each
row is divided by its second-block entry, a rescaling that cannot change
which sets of rows are independent, so the single anchor coordinate
\(z_e\) plays the role of his ratio of the two indeterminates on edge
\(e\).

\hypertarget{a.6-proposition-2}{%
\subsection{A.6 Proposition 2}\label{a.6-proposition-2}}

\textbf{Proof.} If a selected row set splits into two forests \(S_1\)
and \(S_2\), each forest's incidence rows have full row rank, so
node-column sets \(C_1\) and \(C_2\) exist with
\(\det A[S_j,C_j]\neq0\). In the square minor of \(H(z)\) taking
first-block columns \(C_1\) and second-block columns \(C_2\), the
allocation that sends \(S_1\) to the weighted block contributes
\(\pm(\prod_{e\in S_1}z_e)\det A[S_1,C_1]\det A[S_2,C_2]\), which is
nonzero, and any other allocation uses a different subset of the
indeterminates and so carries a different monomial. Nothing can cancel
it, so the minor is a nonzero polynomial and the rows are generically
independent. Conversely every row subset \(X\) satisfies
\(\operatorname{rank}H(z)[X,\cdot]\le2\operatorname{rank}A[X,\cdot]=2(M-c(X))\),
so a generically independent set obeys \(|X|\le2(M-c(X))\) for all its
subsets, which by the matroid-union theorem is precisely the criterion
for splitting into two forests. The rank formula is the matroid-union
rank formula for two identical cycle matroids. For the deficit form,
substitute and rewrite the minimum as
\(\max_F\{2(c(F)-1)-|E\setminus F|\}\). Any \(F\) induces the partition
of the charts into the components of \((V,F)\), and every crossing edge
lies in \(E\setminus F\), so that partition's deficit is at least
\(F\)'s value. Conversely a maximizing partition may be taken to have
connected parts, since refining a part into its internally connected
pieces adds parts without adding crossing edges; taking \(F\) to be all
edges internal to such a partition's parts attains the same value.
\(\square\)

Proposition 2 has three parts, and they do not all belong to us. The
first part states that a set of anchor rows is generically independent
exactly when it splits into two forests. That part is Whiteley's Theorem
1 at \(k=2\): he proves that a set of edges is independent in the
general \(k\)-frame if and only if it is the union of \(k\) forests,
that is \(k\) independent sets in the cycle matroid (1988, p.~239). His
Corollary 3 restates this as the count \(|E'|\le k|V'|-k\) on every
nonempty edge subset, and his proof of it uses the rank function
\(k(|V'|-c(E'))\) that our formula also needs (1988, p.~240). So the
two-copy matroid structure is his, and we claim none of it. The second
part, the generic rank of a design that falls short, is the standard
matroid-union rank formula rather than a result of his paper, which
characterizes independence and the covering count but states no rank
formula. The third part restates the standard matroid-union rank formula
as a partition deficit for this calibration design, which turns that
rank into a number a designer can act on, namely how many affine
directions a given split of the charts leaves unidentified; we claim no
new matroid-union theorem. The paper's own contribution is therefore the
deficit reading as a design quantity and the use of the whole structure
for typed cross-method calibration, together with the contrast between
generic and realized anchor geometry below. That a design short of Tan
et al.'s merging condition (groups of charts that share at least two
anchors merged, repeatedly, into one), \(K_4\) included, fails for some
anchor values already follows from the proof of their Theorem 3.4 (2021,
p.~27073; arXiv:2110.12607, Appendix A.1), whose failing instance is the
additive family \(z_{ij}=u_i+u_j\) used for \(K_4\) in Section 3.6; the
non-additive exceptional assignment, the identification of the
exceptional set with the determinant-zero variety, and the application
of the standard nonzero-polynomial argument that gives the measure-zero
statement of Proposition 3 are ours.

\hypertarget{a.7-corollary-2}{%
\subsection{A.7 Corollary 2}\label{a.7-corollary-2}}

\textbf{Proof.} If the inequality holds, deleting one edge lowers any
partition's crossing count by at most one, leaving
\(\operatorname{cr}(\mathcal P)\ge2(|\mathcal P|-1)\) for every
partition, which by Proposition 2 and Nash-Williams--Tutte gives two
edge-disjoint spanning trees in the reduced graph. Conversely, if the
inequality fails, some partition with at least two parts has
\(\operatorname{cr}(\mathcal P)\le2(|\mathcal P|-1)\). If the count is
below the bound, the design already lacks two edge-disjoint spanning
trees and stays unidentified after any deletion. If it equals the bound,
at least one edge crosses the partition, and deleting that edge drops
the count below the bound, so the reduced graph cannot contain two
edge-disjoint spanning trees. \(\square\)

\hypertarget{a.8-the-angle-bracket-for-the-section-4-statistic}{%
\subsection{A.8 The angle bracket for the Section 4
statistic}\label{a.8-the-angle-bracket-for-the-section-4-statistic}}

Section 4 uses the statistic \(\Delta_e\) of Section 3.4 with \(W_e\)
the identity and \(\rho_e\) a Deming slope through the origin. Before
noise correction, the statistic then depends on the two residual
profiles only through their lengths, the angle \(\theta\) between them,
and the fitted ratio. This appendix shows that it lies between
\(\sin(\theta/2)\) and \(\tan(\theta/2)\).

Write \(a=\|y_u\|\), \(b=\|y_v\|\), \(c=\cos\theta\), and
\(s=\rho_ea/b\), the fitted ratio measured against the ratio of the two
lengths. Expanding the numerator gives

\[
\Delta_e^2=f(s)=\frac{1+s^2-2sc}{(1+s)^2}.
\]

In words, the squared statistic depends only on the relative scale \(s\)
and the angle.

\textbf{Lower bound, for any positive ratio.} The derivative is
\(f'(s)=2(1+c)(s-1)/(1+s)^3\), so for any angle short of \(180\) degrees
\(f\) falls until \(s=1\) and rises after it. Its minimum is
\(f(1)=(1-c)/2=\sin^2(\theta/2)\). Hence \(\Delta_e\ge\sin(\theta/2)\)
whatever positive ratio is fitted, Deming or not.

\textbf{Upper bound, for the Deming ratio.} Suppose the angle is acute,
\(0<c<1\); every edge that passes gate 3 has a positive shared
cross-product (the inner product of the two profiles) and so an acute
angle. The difference \(f(s)-\tan^2(\theta/2)\), with
\(\tan^2(\theta/2)=(1-c)/(1+c)\), equals \(2(cs-1)(s-c)\) divided by the
positive quantity \((1+c)(1+s)^2\). So \(f(s)\le\tan^2(\theta/2)\)
exactly when \(c\le s\le1/c\). With error-variance ratio \(\lambda>0\)
(the \(v\) block's error variance over the \(u\) block's), the Deming
slope through the origin is
\(\rho_e=\{b^2-\lambda a^2+[(b^2-\lambda a^2)^2+4\lambda a^2b^2c^2]^{1/2}\}/(2abc)\).
It lies between the least-squares slope of \(y_v\) on \(y_u\), \(bc/a\),
which is \(s=c\), and the reciprocal of the least-squares slope of
\(y_u\) on \(y_v\), \(b/(ac)\), which is \(s=1/c\), for every
\(\lambda>0\). To see this, write \(t=\lambda a^2/b^2\), so that
\(s=[1-t+\{(1-t)^2+4tc^2\}^{1/2}]/(2c)\). Then \(s\le1/c\) reduces to
\((1-t)^2+4tc^2\le(1+t)^2\), that is \(c^2\le1\), and \(s\ge c\)
reduces, after squaring where the right side is positive, to
\((2c^2-1)^2\le1\); both hold for \(0<c<1\). The fitted \(s\) therefore
lies in \([c,1/c]\), and \(\Delta_e\le\tan(\theta/2)\). The bound is
approached only as \(\lambda\) tends to zero or infinity, where the
Deming slope becomes one of the two least-squares slopes.

\textbf{Consequence for the margin.} Together,
\(\sin(\theta/2)\le\Delta_e\le\tan(\theta/2)\) for an acute angle. For
the margin 0.20 the lower bound gives a necessary condition: an
uncorrected statistic below 0.20 requires \(\theta<2\arcsin0.2\), about
23.07 degrees, a squared cosine above about 0.846. The upper bound gives
a sufficient condition: \(\theta<2\arctan0.2\), about 22.62 degrees,
guarantees it. These two angles lie less than half a degree apart, and
they give Section 4's ``about 23 degrees'' and its squared correlation
of ``about 0.85''. Because the residual profiles have mean zero,
\(\cos\theta\) is their correlation.

\textbf{Check.} Wolfram Mathematica (\texttt{Resolve} over the reals,
for all \(a,b>0\), \(0<c<1\), \(\lambda>0\)) confirmed the factorisation
of \(f'\), the minimum at \(s=1\), the equivalence of
\(f(s)\le\tan^2(\theta/2)\) with \(c\le s\le1/c\), the position of the
Deming slope with both reductions, and the combined bracket. Four
negative controls returned False, as they must: the claim
\(\Delta_e\le\sin(\theta/2)\) for every ratio, the upper bound for
obtuse angles, the upper bound at a ratio outside \([c,1/c]\), and the
claim \(s\ge1\) for every \(\lambda\).

\hypertarget{appendix-b-computational-checks}{%
\section{Appendix B: Computational
checks}\label{appendix-b-computational-checks}}

\hypertarget{b.1-graph-verification}{%
\subsection{B.1 Graph verification}\label{b.1-graph-verification}}

The incidence nullity identity was checked in 44,776 cases: every
labeled simple graph through six vertices, 33,867 in all, and 10,909
seeded random weighted graphs with exact zero sensitivities. The
exhaustive portion covers every simple graph a design on six or fewer
blocks could have, though not the multigraphs that repeated anchors
between one pair would create; the weighted cases stress-test
sensitivity to exact zeros. No mismatch occurred. The weighted cases
were regenerated for this version from the recorded seed 20261012
(\texttt{anc/p50\_graph\_and\_s1\_checks.py}, which also computes the S1
ranks of Appendix B.4). These checks validate the implementation's code
paths rather than adding evidence for the proof.

\hypertarget{b.2-affine-gauge-simulation}{%
\subsection{B.2 Affine-gauge
simulation}\label{b.2-affine-gauge-simulation}}

The frozen generator uses a \(7\times7\) grid on

\[
[-0.75,0.75]^2.
\]

Its latent residual profile is

\[
P\left\{
\sin(2.2x)\cos(1.7y)
+
0.55(x^2-y^2)
+
0.25xy
\right\},
\]

where \(P\) projects off

\[
[\mathbf1,x,y];
\]

the residual is rescaled to a fixed signal standard deviation of 1.055.
That value is a frozen simulation normalization retained for
reproducibility, not an empirical estimate.

With a true scale ratio of 1.4, independent observer affine-plane
coefficients with standard deviation 0.75, point noise with standard
deviation 0.08, and 100,000 replicates, master seed 20260916 gave the
results in Table 4 (script \texttt{anc/p50\_gauge\_sim.py}). The columns
give the mean estimated ratio, its bias against the true 1.4, the
root-mean-square error (RMSE), and the proportion of replicates within
10\% of 1.4. These frozen settings define an illustrative regime with
sizeable affine-plane nuisance and much smaller point noise; they are
not estimates from an artwork dataset. The replicate count is large
enough that Monte Carlo error is negligible relative to the differences
between the two procedures in the table.

\textbf{Table 4. Affine-gauge simulation: recovery of a true scale ratio
of 1.4 with and without projecting out the declared nuisance basis.}

\begin{longtable}[]{@{}lrrrr@{}}
\toprule
Estimation procedure & Mean ratio & Bias & RMSE & Within 10\% \\
\midrule
\endhead
Mean-center only & 1.148 & -0.252 & 0.337 & 0.379 \\
Project out \([\mathbf1,x,y]\) & 1.393 & -0.007 & 0.020 & 1.000 \\
\bottomrule
\end{longtable}

The small remaining attenuation is expected from error in the reference
axis; a future empirical application should combine gauge reduction with
errors-in-variables or other joint measurement-error inference. Section
4 applies gauge reduction to one published participant-level dataset and
estimates each gain ratio by errors-in-variables regression.

\hypertarget{b.3-correlated-error-simulation}{%
\subsection{B.3 Correlated-error
simulation}\label{b.3-correlated-error-simulation}}

In 200,000 four-style null simulations per cell, the log-ratio
covariance was equicorrelated with unit marginal variances. Rejection
rates of the generalized least-squares (GLS) test with the correct
covariance were .0502, .0503, and .0504 for correlations
\((-.30,0,.60)\). Replacing the covariance by the identity gave .1122,
.0503, and .0002. Thus assuming independence was anti-conservative under
negative correlation and strongly conservative under positive
correlation in this generator. The sign and magnitude are
design-dependent; estimating covariance is the rule. The original run's
seeds were not preserved with its code; the replacement script that
reproduces these rates (Data and code availability) uses master seed
2026100902.

\hypertarget{b.4-frozen-mixed-method-simulations}{%
\subsection{B.4 Frozen mixed-method
simulations}\label{b.4-frozen-mixed-method-simulations}}

S1 checks realized affine-incidence rank with exact integer arithmetic.
On charts \(0,1,2,3\), the doubled path uses two anchors on each edge:
\(z=0\) and \(1\) on edge \(01\), \(0\) and \(1\) on \(12\), and \(0\)
and \(4\) on \(23\). The selected gauge-fixed full minor has determinant
4. The well-spread \(K_4\), with anchors \((-2,-2,0,2,-2,3)\) on edges
\((01,02,03,12,13,23)\), has determinant \(-40\). Both determinants are
nonzero, which is their whole role here: each certifies that the
selected gauge-fixed minor has full rank. Their magnitudes and signs are
artifacts of the chosen orientation and column selection and carry no
substantive meaning. The three-edge tree has calibration rank three; the
doubled-path and well-spread \(K_4\) designs each reach the full
calibration rank six; the constructed \(K_4\) assignment

\[
z_{ij}=u_i+u_j,
\qquad
u=(0,1,3,7),
\]

has rank five. A two-component pair of doubled edges has rank four, the
componentwise target

\[
2(M-c)=4,
\]

and remains two dimensions short of one global affine coordinate. These
checks establish algebraic rank, not finite-sample coverage or model
fit.

S2, the joint simulation of two continuous charts and one
probabilistic-comparison chart, uses 12 equally spaced latent scores
from \(-1.65\) to \(1.65\). Continuous chart A observes the even-indexed
stimuli and fixes the global affine gauge; chart B observes the
odd-indexed stimuli as \(y_B=1.55z+.25\) plus noise, a forward
generating slope of 1.55 and offset .25, so its Section 3.6 calibration
slope is \(1/1.55\). A Bradley--Terry chart with discrimination 1
observes all 66 pairs. In Bradley and Terry's (1952) model, stimulus
\(i\) is preferred to stimulus \(j\) with probability
\(\pi_i/(\pi_i+\pi_j)\) (p.~325); with \(\pi_i\) proportional to
\(e^{z_i}\), that probability is Section 5.1's model with a logistic
link, discrimination 1, no order bias and no lapse. Both continuous
charts carry Gaussian measurement noise with standard deviation .10. The
joint likelihood, Gaussian for the two charts and binomial for the
comparisons, estimates the latent scores, chart B's positive slope and
offset, and the comparison discrimination; log-slope is searched over
\([-3,3]\).

Each trial-count cell uses 500 repetitions from its own stream, spawned
from master seed 2026100903 with NumPy's SeedSequence. The three trial
counts create low-, medium-, and higher-information versions of the same
identified model, so precision changes while algebraic identifiability
does not. With 5, 20, and 80 trials per pair, the RMSE of chart B's
forward slope (true value 1.55) was .303, .155, and .105; the mean
number of stimulus pairs ranked in the wrong order was 1.334, .256, and
.016. Every profile interval had two finite endpoints in the declared
log-slope range. Empirical coverage was .936, .948, and .936; with 500
repetitions the Monte Carlo standard error is about .010, so each lies
within two standard errors of the nominal 95\% level in this generator.
The median width was 1.191, .633, and .402. These values depend strongly
on the continuous charts' noise standard deviation, which sets how much
of the slope information comes from charts A and B rather than from the
comparisons.

These results describe this generator and fitter. Collapsing the trials
to a final ordering, while keeping the continuous-chart measurements,
removes point identification of chart B's slope in this disjoint-chart
design: small positive affine changes of chart B preserve every ordering
inequality. The full trial counts do identify it, through the comparison
probabilities. The complete comparison graph is connected, and the
population triangle-cycle discrepancy, computed from the linear
predictor, is below

\[
5\times10^{-16},
\]

that is, zero to floating-point precision. This is a consistency check
on the generator, not an empirical goodness-of-fit result.

Figure 4 shows the prespecified first-repetition profile in each
trial-count cell.

\includegraphics{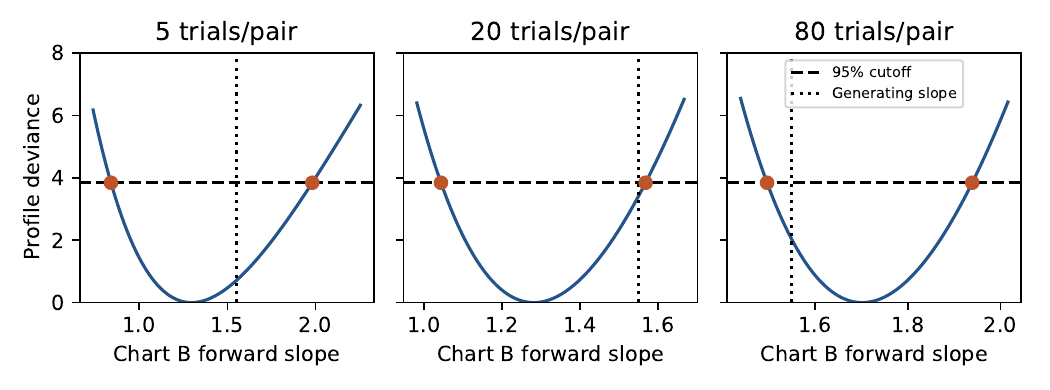}

Figure 4. S2 likelihood profiles: profile deviance for the first
frozen-seed repetition at each trial count. Dots at the dashed 95\%
cutoff mark the computed interval endpoints; the dotted line marks the
generating slope, 1.55. The curves illustrate one sample per cell, while
coverage is calculated from all 500 repetitions. Source:
\texttt{anc/p50\_s2\_mixed\_method\_sim.py}.

\begin{quote}
\emph{Two gaze at painted hills, yet each may see}\\
\emph{A different depth within the same design.}\\
\emph{Connected paths can make their scales agree,}\\
\emph{But shared coordinates take more to align.}
\end{quote}

\end{document}